\documentclass[twocolumn]{aa}
\usepackage{booktabs}
\usepackage{graphicx}
\usepackage{amsmath,amsfonts,amssymb}
\usepackage{txfonts}
\usepackage{color}
\usepackage{natbib}
\usepackage{float}
\usepackage{dblfloatfix}
\usepackage{afterpage}
\usepackage{ifthen}
\usepackage[morefloats=12]{morefloats}
\usepackage{placeins}
\usepackage{multicol}
\bibpunct{(}{)}{;}{a}{}{,}
\usepackage[switch]{lineno}
\definecolor{linkcolor}{rgb}{0.6,0,0}
\definecolor{citecolor}{rgb}{0,0,0.75}
\definecolor{urlcolor}{rgb}{0.12,0.46,0.7}
\usepackage[breaklinks, color links, urlcolor=urlcolor,
    linkcolor=linkcolor,citecolor=citecolor,pdfencoding=auto]{hyperref}
\hypersetup{linktocpage}
\usepackage{bold-extra}
\usepackage{algpseudocode}
\usepackage{comment}

\usepackage{lipsum}

\providecommand{\sorthelp}[1]{}

\begin{document}

\title{Evidence for cloud-to-cloud variations in the ratio of polarized thermal dust emission to starlight polarization}
\author{\small N.~Mehandiratta\inst{1, 2}\thanks{\url{nidhi@saao.ac.za}}
\and
G.~Panopoulou\inst{3}
\and
E.~Gjerløw\inst{4}
\and
V.~Pelgrims\inst{5,6}
\and
K.~Tassis\inst{7,8}
\and
D. Blinov\inst{7,8}
\and
B. Hensley\inst{14}
\and
J. A. Kypriotakis\inst{8}
\and
S. Maharana\inst{9}
\and
N. Mandarakas\inst{10}
\and
V. Pavlidou\inst{7,8}
\and
S. B. Potter\inst{1,11}
\and
A. N. Ramaprakash\inst{12,8,13}
\and
R. Skalidis\inst{15}\fnmsep\thanks{Hubble Fellow}
\and
N. Uppal\inst{8}
}
\institute{\small
South African Astronomical Observatory, PO Box 9, Observatory 7935, Cape Town, South Africa\goodbreak
\and
Department of Astronomy, University of Cape Town, Private Bag, Rondebosch 7701, Cape Town, South Africa\goodbreak
\and
Department of Space, Earth and Environment, Chalmers University of Technology, 412 93, G\"{o}teborg, Sweden\goodbreak
\and
Institute of Theoretical Astrophysics, University of Oslo, Blindern, Oslo, Norway\goodbreak
\and
Universit{\'e} Libre de Bruxelles, Science Faculty CP230, B-1050 Brussels, Belgium
\and
Universit{\'e} Paris-Saclay, Universit{\'e} Paris Cit{\'e}, CEA, CNRS, AIM, F-91191 Gif-sur-Yvette, France
\and
Department of Physics, University of Crete, Vasilika Vouton, 70013, Heraklion, Greece
\and
Institute of Astrophysics, Foundation for Research and Technology – Hellas, 100 Nikolaou Plastira, Vassilika Vouton, 70013, Heraklion, Greece
\and
University of Oxford, Denys Wilkinson Building, Department of Physics, Oxford, OX1 3RH, UK
\and
Aix Marseille Univ., CNRS, CNES, LAM, Marseille, France
\and
Department of Physics, University of Johannesburg, PO Box 524, Auckland Park, 2006, South Africa
\and
Inter-University Centre for Astronomy and Astrophysics, Post bag 4, Ganeshkhind, Pune, 411007, India
\and
Cahill Center for Astronomy and Astrophysics, California Institute of Technology, Pasadena, CA, 91125, USA
\and
Jet Propulsion Laboratory, California Institute of Technology, 4800 Oak Grove Drive, Pasadena, CA 91109-8099,
USA
\and
TAPIR, California Institute of Technology, MC 350-17, Pasadena, CA 91125, USA
}

\authorrunning{Mehandiratta et al.}
\titlerunning{Evidence for cloud-to-cloud variations in $R_{P/p}$}

\abstract{The correlation between optical starlight polarization and polarized thermal dust emission can be used to infer intrinsic dust properties. This correlation is quantified by the ratio $R_{P/p}$, which has been measured by the Planck Collaboration to be $5.42 \pm 0.05\,\mathrm{MJy\,sr^{-1}}$ at $353\,\mathrm{GHz}$ when averaged over large areas of the sky.
  We investigate this correlation 
  using newly published stellar polarimetric data densely sampling a continuous sky region of about four square degrees at intermediate Galactic latitude.
  We combine RoboPol optical polarization measurements for 1,430 stars with submillimeter data from the \textit{Planck} satellite at 353\,GHz. 
  We perform linear fits between the {\textit{Planck}} ($Q_s$, $U_s$) and optical ($q_v$, $u_v$) Stokes parameters, taking into account the differences in resolution between the two datasets as well as the distribution of clouds along the line of sight.
  We find that in this region of the sky the $R_{P/p}$ value is \(3.67 \pm 0.05\)\,MJy\,sr$^{-1}$, indicating a significantly shallower slope than that found previously using different stellar samples.
  We also find significant differences in the fitted slopes when fitting the \(Q_s\)--\(q_v\) and \(U_s\)--\(u_v\) data separately. We explore two explanations using mock data: miscalibration of polarization angle and variations in $R_{P/p}$ along the line of sight due to multiple clouds. 
  We show that the former can produce differences in the correlations of \(Q_s\)--\(q_v\) and \(U_s\)--\(u_v\), but large miscalibration angles would be needed to reproduce the magnitude of the observed differences. Our simulations favor the interpretation that $R_{P/p}$ differs between the two dominant clouds that overlap on the sky in this region. The difference in \( R_{P/p} \) suggests that the two clouds may have distinct dust polarimetric properties. With knowledge from the tomographic decomposition of the stellar polarization, we find that one cloud appears to dominate the correlation of \(U_s\)--\(u_v\), while both clouds contribute to the correlation of the \(Q_s\)--\(q_v\) data.}
\keywords{polarization -- ISM: dust, extinction -- magnetic fields -- Cosmology: observations -- cosmic microwave background -- diffuse radiation -- Galaxy: general}


\maketitle

\section{Introduction}
Stellar polarization \,(due to dichroic extinction) and polarized thermal emission from dust grains are manifestations of the alignment of aspherical dust grains with the ambient Galactic magnetic field \citep{Davis1951A, Lazarian2007, Andersson2015}. These polarized observables provide crucial insights into the structure of interstellar magnetic fields \citep{Pattle2023} and the physical properties of dust grains \citep{Draine2003ARA&A,Hensley2021}. As such, understanding these phenomena is key to both Galactic astrophysics and the precise removal of foreground contamination in cosmological studies \citep{Page2007,Planck2015, Planck2015stars}. 


The correlation between the plane-of-sky orientation of polarized dust emission and polarized starlight has been studied in a variety of regions \citep[e.g.][]{Soler2016}. Comparatively less attention has been given to the correlation between the polarization amplitudes of these signals, despite a close link having been anticipated  \citep[e.g.][and references therein]{Martin2007}. \citet{Planck2015stars} combined archival measurements of optical stellar polarization with polarized dust emission data at 353 GHz and demonstrated  that the Stokes parameters in absorption and emission are tightly correlated. Their correlation was quantified by the ratio of the polarized intensity of the emission ($P$) to the optical polarization fraction ($p$), $R_{P/p} = P/p$. This ratio was found to have a value of approximately $5.4 \pm 0.5\,\mathrm{MJy\,sr^{-1}}$ at 353~GHz \citep{Planck2015stars}. A later study \citep{Planck2018} focusing on a sample of stars at high Galactic latitudes \citep{berdyugin2001a,berdyugin2002, berdyugin2001b,berdyugin2004,berdyugin2014} found a similar value of $R_{P/p}$ of $5.42 \pm 0.05\,\mathrm{MJy\,sr^{-1}}$. The value of $R_{P/p}$ was found to be largely independent of column density (up to 10\% variations), dust temperature, resolution and Galactic latitude \citep{Planck2015stars,Planck2018}. 
This ratio has proven to be a powerful tracer of dust grain physics, providing constraints for parameters in dust models such as grain alignment efficiency, grain shape, and porosity \citep{Guillet2018, Draine2021, Hensley2023}.

Beyond its role in dust astrophysics, this ratio could be used as a potential tool for modeling polarized foregrounds in Cosmic Microwave Background (CMB) studies. However, for it to serve as a reliable predictor, its homogeneity across the sky must be established. Recent studies have reported spatial variations in the ratio in a small region of the diffuse ISM \citep{Panopoulou2019}, and within a molecular cloud \citep{Santos2017}, suggesting that the assumption of a uniform correlation may not hold. These variations may be driven by changes in dust grain properties (e.g. \citealt{Guillet2018, Draine2021}) or by line-of-sight effects, where stars and dust emission trace different column densities in the three-dimensional interstellar medium (ISM) \citep{Skalidis2019}. Moreover, the wavelength dependence of stellar polarization (\citealt{Serkowski1975}) and the angular smoothing scale of the maps used for comparison can both influence the inferred ratio, adding further complexity to its interpretation.


In this work, we investigate the possible variation of the ratio of polarized thermal dust emission at 353\,GHz to optical starlight polarization using starlight polarization data within a continuous and extended region of approximately four square degrees.
We combine starlight polarization measurements with \textit{Planck} 353\,GHz polarization maps
(\texttt{Nside = 2048}, $\sim5$~arcmin) and pixel-average the data to a common
\texttt{Nside = 256} grid ($\sim14$~arcmin) for the comparison.
 This region offers several advantages for our study, as it is densely sampled by stellar polarization data and has a well-characterized three-dimensional magnetized ISM structure \citep{Pelgrims2024}. The latter was obtained from a tomographic decomposition of the stellar polarization \citep{Pelgrims2023}, as a demonstration of the capabilities of the PASIPHAE survey \citep{tassis2018pasiphaehighgalacticlatitudehighaccuracyoptopolarimetric}.

The structure of the paper is as follows. Section~\ref{sec:data} describes the data sets used in our analysis. Section~\ref{sec:method} outlines the methodology. Section~\ref{sec:results} presents the main results. Section~\ref{discussion} discusses the physical implications and Sect.~\ref{summary} concludes with a summary of our findings.

    

\section{Data}
\label{sec:data}

\subsection{{\textit{Planck}} data products}

\noindent We used full-sky submillimeter polarization data obtained by the \textit{Planck} satellite, specifically from the fourth data release of the \textit{Planck} collaboration (PR4) \citep{planck2020_data_processing}. We used the single-frequency maps of Stokes $Q$, $U$ at 353\,GHz downloaded from the \textit{Planck} Legacy Archive\footnote{\url{http://pla.esac.esa.int/pla/\#maps};\newline\nolinkurl{HFI_SkyMap_353-BPassCorrected-field-IQU_2048_R4.00_full.fits}}. Galactic thermal dust emission dominates the polarized sky at 353 GHz \citep[contributions from CMB polarization and polarized Cosmic Infrared Background (CIB) anisotropies are subdominant][]{Planck2018results}.

The dataset consists of the Stokes parameters $I_s$, $Q_s$, and $U_s$ of the linear polarization of the dust emission and contains the per-pixel block-diagonal noise-covariance matrices between these observables. The data are provided at native resolution of the \textit{Planck} mission, which adopts a HEALPix\footnote{\url{https://healpy.readthedocs.io/en/latest/}} tessellation of the sphere with a resolution parameter of $N_{\mathrm{side}} = 2048$ \citep{Gorski2005,Zonca2019}.
Since data are provided in units of K$_\mathrm{CMB}$, we multiplied them by a factor of 287.45 to convert them into MJy sr$^{-1}$. This factor converts from thermodynamic temperature to specific intensity at 353~GHz, and was obtained from the \textit{Planck} unit conversion and color correction tables, which account for the bandpass response of the High Frequency Instrument (HFI) and the non-linear relationship between temperature and intensity \citep{Planck2013A9}.

\subsection{Optical starlight polarization catalog}
\label{sec:optical data}
We used the optical polarization catalog of \citet{Pelgrims2024}.
The catalog contains starlight polarization measurements and the corresponding uncertainties for 1698 stars over an area of approximately four square degrees
\((\approx 3^\circ \times 1.5^\circ)\) centered on Galactic coordinates \((l, b) \approx (103.5^\circ, 22.25^\circ)\)
where $l$ and $b$ are the Galactic longitude and latitude, respectively. Measurements were made in the Johnson-Cousins R band using the RoboPol optical polarimeter (\citealt{Ramaprakash2019}) mounted at the 1.3-m telescope at the Skinakas Observatory\footnote{\url{https://skinakas.physics.uoc.gr}}. 
The stellar polarization data are measured in the IAU convention (polarization angle measured from zero at the north and increasing towards the east). These measurements are provided in Equatorial coordinates, whereas the \textit{Planck} data are in Galactic coordinates. For consistency, we transformed the stellar data to Galactic coordinates following the procedure described in \citet{Pelgrims2024}.

We use \textit{Gaia} EDR3 source identifier of stars given in the catalog to obtain distances from \citet{Bailer-Jones2021}.
There are 1675 stars with probabilistic distance estimates from \citet{Bailer-Jones2021}. We use the photogeometric distance estimates from their work, which take into account both photometry and parallax information. From this set, we consider only the 1,430 stars flagged as  `ISM probe' in the catalog compiled by \citet{Pelgrims2024},
ensuring that the sample traces the dusty magnetized ISM and excludes intrinsically polarized stars (see \citealt{Pelgrims2024} for details).

\label{intrinsic_stars}

\subsubsection{Star selection based on distance} 
\label{sec:distance_cuts}

As opposed to polarized dust emission, which probes the full column density up to infinity, starlight polarization only traces the dusty magnetized ISM up to the distances of the stars. To minimize any discrepancy between the amount of dust traced in absorption and in emission, we choose to select stars that likely trace most of the dust column, taking into account the knowledge of the 3D distribution of polarizing dust along the line of sight as reconstructed by \citet{Pelgrims2024}. This approach differs from the one adopted in \citet{Planck2015stars,Planck2018} where the stellar polarization data were artificially corrected for the missing dust column by comparing the stellar extinction to the total extinction. This latter approach has limitations due to the available stellar-extinction maps and disregards possible depolarization effects due to variation of the plane-of-sky component of the magnetic field with distance.

\citet{Pelgrims2024} found that two clouds dominate the stellar polarization signal in this region of the sky.
The first cloud is located at $\sim$ 58\,pc (with considerable uncertainty as posterior distributions on its distance peak between 30 and 170 pc) and contributes, on average, a polarization fraction of $0.19\%$ with an average polarization angle of $80^\circ$. The second, more distant, cloud is located at 374~pc (with estimates ranging between 354 - 448 pc). It contributes more significantly to the polarization, with a median polarization fraction of $1.45\%$ and an average polarization angle of $40^\circ$.
The two clouds cover the entire sky region considered in this work.
To optimize the correlation between dust polarization in emission and absorption (starlight polarization), we selected stars with distances larger than 450~pc. This threshold corresponds to a distance beyond the 95th percentile of the cloud's posterior distribution, which peaks at \( 374 \,\mathrm{pc} \), thus ensuring that the selected stars lie behind the cloud with high confidence.

More distant clouds were found towards distinct sightlines, at distances ranging from 1600~pc to 2400~pc in \citealt{Pelgrims2024}. These clouds can induce non-negligible polarization fractions (up to 0.58\%) but only in few isolated regions of the map, each of size of approximately $10\arcmin$.
These very distant clouds affect only a small fraction of our stellar sample and should therefore only be considered as an additional source of noise in our analysis.
To validate this assumption and check the robustness of our analysis with regard to these distant clouds, we also repeat our analysis with two other distance cuts, one demanding $d > 2$~kpc and the other $d \in [450,\,2000]$~pc.


To apply our distance cut, taking into account uncertainties on stellar distances, we used the lower and upper bounds of the photogeometric distance estimates from \citet{Bailer-Jones2021}.
Specifically, stars in the 450--2000~pc range are selected by requiring the lower bound (\texttt{r\_lo\_photogeo}) to be greater than 450~pc and the upper bound (\texttt{r\_hi\_photogeo}) to be less than 2000~pc, yielding a final sample of 929 stars. For stars confidently beyond certain thresholds, we rely solely on the lower bound: \texttt{r\_lo\_photogeo} $>$ 450~pc results in 1200 stars, while \texttt{r\_lo\_photogeo} $>$ 2000~pc results in 228 stars.

\section{Methods}
\label{sec:method}
\cite{Planck2018} used starlight polarization data from a compilation of catalogs (\citealt{berdyugin2001b, berdyugin2004, berdyugin2014}; \citealt{berdyugin2001a, berdyugin2002})
to compare against the Stokes parameters of the polarized dust emission. Unlike optical starlight data, which probes narrow lines of sight, {\textit{Planck}} measurements average over larger beams ($\gtrsim$5~arcmin), leading to potential depolarization effects. To account for this, \citet{Planck2018} applied a correction factor to the Stokes parameters and associated uncertainties based on an analytical model of beam depolarization. 

In our analysis, we do not apply this type of correction. Instead, we pixelize the optical and {\textit{Planck}} data using the HEALPix scheme.
 The motivation for this pixelization is two-fold: (i) to improve the signal-to-noise ratio by averaging over several measurements per pixel and (ii) to ensure that each data point used in our comparison is as independent as possible. This is particularly important because the sampling of starlight polarization is often dense, with multiple stars falling within even a high-resolution $N_{\rm{side}} = 2048$ pixel. After experimenting with different resolutions, we adopt $N_{\rm{side}}=256$ as a compromise between statistical independence and data quality. 
For example, approximately 80\% of the pixels at $N_{\rm{side}} = 256$ contain at least five stars, in contrast to $N_{\rm{side}} = 512$, where 65\% of the pixels contain fewer than five stars. Therefore, we expect to obtain more reliable statistical estimates at $N_{\rm{side}} = 256$ and reduce sensitivity to beam mismatch effects between \textit{Planck} and stellar polarization data \citep{Planck2018}.

\subsection{ Pixelization}
\label{sec:pixelization}
Here, we describe the pixelization scheme used for both datasets. Although the method is generic, we adopt the convention that lowercase Stokes parameters ($q$, $u$) denote fractional optical starlight polarization, whereas uppercase Stokes parameters ($Q$, $U$) refer to \textit{Planck} thermal dust emission at 353\,GHz, expressed in $\mathrm{MJy\,sr^{-1}}$.

\begin{figure*}[ht!]
    \centering
    \includegraphics[scale = 0.6]{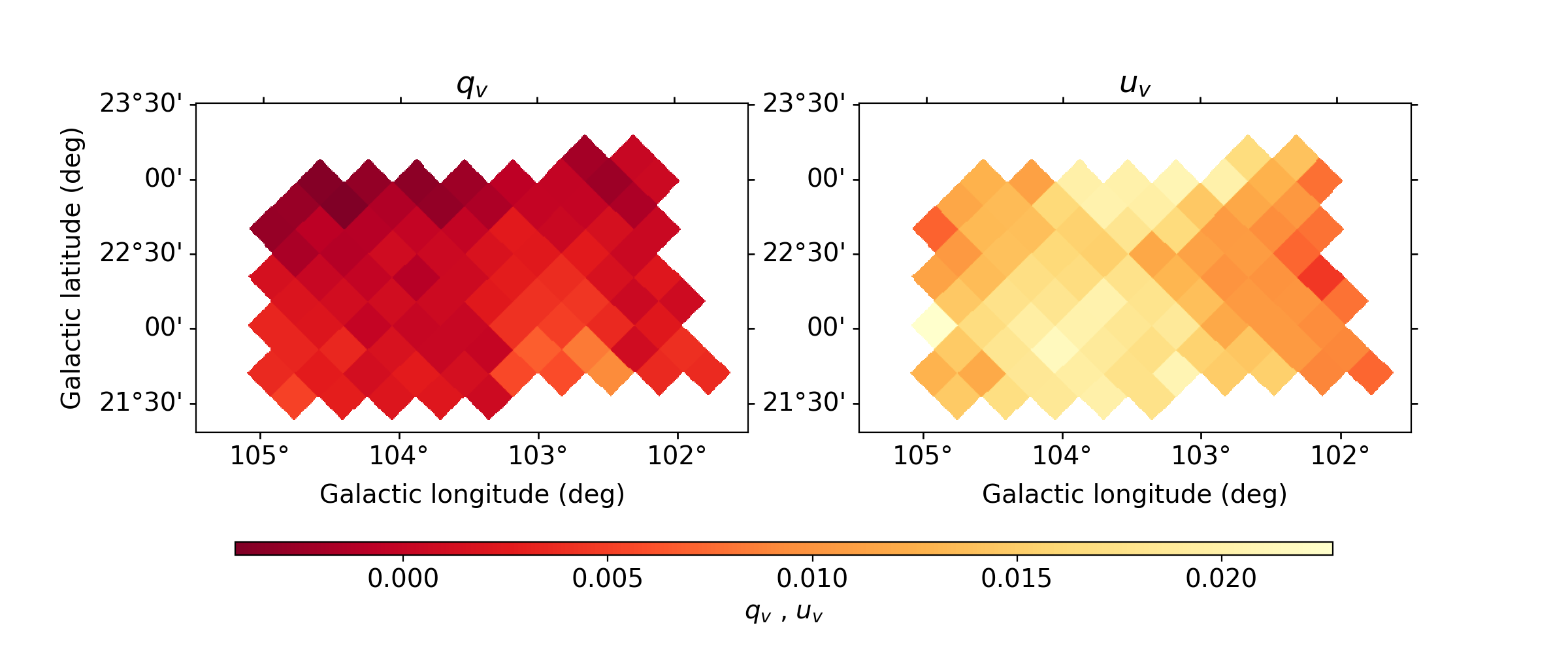}
%
    \includegraphics[scale = 0.6]{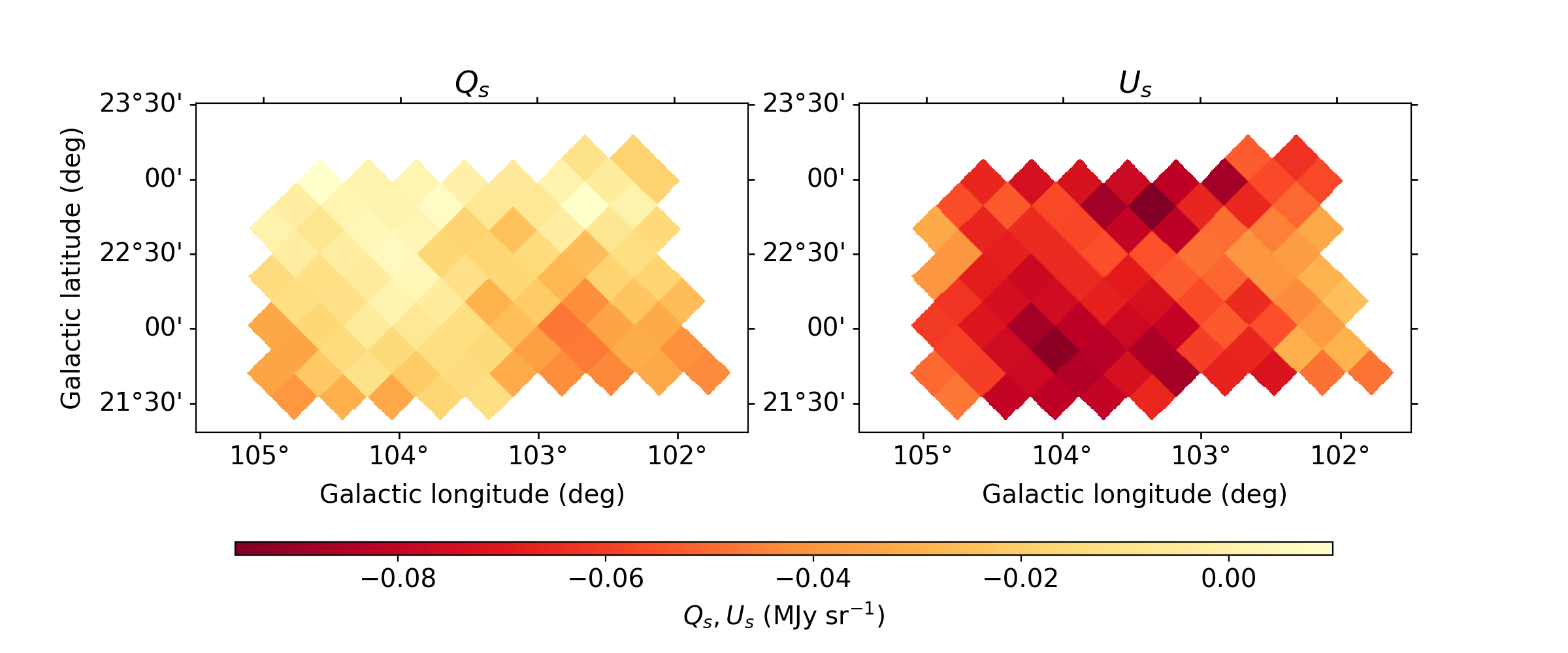}
    \caption{ 
    Gnomonic projections of the pixelized Stokes parameter data at optical (top) and submm (bottom) at $N_{\rm{Side}}$ = 256 and distance cut > 450~pc. Top left: $q_{\rm{v}}$, Top right: $u_{\rm{v}}$, Bottom left: $Q_S$, bottom right: $U_S$. The Stokes parameters follow the IAU convention. Pixels with no optical data (outside of the observed sky region) are colored white.}
     \label{fig:gnomonic}
\end{figure*}

\subsubsection{RoboPol Data}
\label{sec:pixelization_rbpl}

The stellar polarization measurements are pixelized to create a HEALPix map at \texttt{$N_{\rm{side}}$} = 256 (pixel size of $\approx 14 \arcmin$) as follows.

Suppose we have a total of \( N \) star measurements that can be distributed, according to their coordinates, across $N_{\rm{pix}}$ pixels. 
The measured polarization $(q,\,u)$ of each star $i$ in the sample comes with uncertainties. In the Equatorial coordinate system there is no covariance between the measurements of $q$ and $u$ from the RoboPol instrument (\citealt{Ramaprakash2019, Blinov2023}). Therefore, the noise covariance matrix takes a diagonal form:
\begin{equation}
\boldsymbol{C}_{qu, i} = \begin{bmatrix} \sigma_{q, i}^2 & 0 \\ 0 & \sigma_{u,i}^2 \end{bmatrix} \;,
\end{equation}
which is no longer the case when the data is rotated to the Galactic reference frame (e.g., \citealt{Pelgrims2024}).




We assume that within a given pixel $\rm p$, the ISM induces a polarization to the stars, described by a mean $q_{\rm p}$ and $u_{\rm p}$. We further assume that all variations in the observed stellar polarization arise solely from measurement uncertainties, which are normally distributed. We revisit and assess the validity of this assumption in Sect.~\ref{sec:chisq_method}. Our data model for each star measurement $i$ can then be written as 


\begin{equation}
\label{eq:data_model}
\mathbf{d}_i =\begin{bmatrix}
                q_i \\
                u_i
                \end{bmatrix}
                \sim N\left(\begin{bmatrix}
                        q_{\rm p} \\
                        u_{\rm p}
                        \end{bmatrix}, \boldsymbol{C}_{qu, i}\right) \;,
\end{equation}
where $N(\boldsymbol{\mu}, \boldsymbol{\Sigma})$ represents a random draw from a normal bivariate distribution with mean $\boldsymbol{\mu}$ and covariance $\boldsymbol{\Sigma}$.
We also assume that the measurement uncertainties are independent, so that there is no covariance between measurements.

Given this model, an optimal estimator for the pixel weighted-mean polarization $\left\langle \mathbf{d} \right\rangle_{\rm{p}} \coloneqq (q_{\rm{p}},\,u_{\rm{p}})$ is given by

\begin{equation}
\langle \mathbf{d}\rangle_{\rm p} =\left(\sum_i\boldsymbol{C}_{qu,i}^{-1}\right)^{-1}\sum_i\boldsymbol{C}_{qu,i}^{-1}\mathbf{d}_i \;.
\label{eqref:meanPol}
\end{equation}
The covariance matrix corresponding to the weighted-mean polarization 
$\left\langle \mathbf{d} \right\rangle_{\rm{p}}$ is given by

\begin{equation}
\boldsymbol{C}_{qu,{\rm p}} = \left(\sum_i\boldsymbol{C}_{qu,i}^{-1}\right)^{-1}
\label{eqref:covMeanPol}
\end{equation}
For diagonal covariance matrices
$\boldsymbol{C}_{qu, i}$, this resolves trivially into a set of scalar equations.


In the remainder of this work, we denote the pixel-averaged fractional Stokes parameters derived from the RoboPol data as $q_v$ and $u_v$, where the subscript `$v$' indicates starlight polarization in the visible. These correspond to the quantities $q_{\rm p}$ and $u_{\rm p}$ defined above. The associated uncertainties on these parameters are denoted as $\sigma^2_{q_v}$ and $\sigma^2_{u_v}$, obtained from the diagonal elements of the pixel-averaged covariance matrix $C_{qu,{\rm p}}$. Since the stellar measurements are rotated into the Galactic coordinate system,
 we also retain the off-diagonal term, denoted as $\sigma_{qu_v}$, which contributes to the joint error analysis (see Sect.~\ref{sec:fittingproc}).

\subsubsection{{\textit{Planck}} data}
\label{sec:planck_pixelization}

The {\textit{Planck}} data are processed in a manner similar to the RoboPol data. The maps are provided at a native resolution of $N_{\rm{side}}=2048$, where each pixel is treated as an independent measurement of the Stokes parameters \( Q \) and \( U \). 
\noindent Since {\textit{Planck}} follows the HEALPix convention, we convert to the IAU system by multiplying the values of \( U \) by \(-1\) prior to any further processing.

For the comparison with the RoboPol data, we construct a lower-resolution map of $N_{\rm{side}}=256$, where values of the original map are averaged in larger pixels. At this resolution, each large pixel at $N_{\rm{side}}=256$ encompasses 64 smaller pixels from the original $N_{\rm{side}}=2048$ map. These smaller pixels are treated as independent measurements (analogous to individual stars), and their values are grouped using Eqs.~\eqref{eqref:meanPol} and \eqref{eqref:covMeanPol}.


In the remainder of this work, we denote the pixel-averaged \textit{Planck} Stokes parameters as $Q_s$ and $U_s$ ($s$ is used to denote submillimeter polarization). The associated uncertainties on these parameters are denoted as $\sigma^2_{Q_s} = C_{QQ}$ and $\sigma^2_{U_s} = C_{UU}$, which are the diagonal elements of the pixel-averaged covariance matrix. The off-diagonal term of this matrix is denoted as $C_{QU}$.
\\

\color{black}



\color{black}

\subsection{Comparison of dust polarization in emission and absorption}

To compare the \textit{Planck} data with starlight polarization, we aim to measure the polarization ratio \(R_{P/p}\), which in principle could be estimated by correlating the polarized intensity in emission with the polarization fraction in extinction. However, both the polarized intensity and the polarization fraction are non-linear combinations of the Stokes parameters and are positively biased in the presence of noise. 
In contrast, the Stokes parameters have Gaussian noise properties. Following \citet{Planck2015stars}, we therefore quantify the emission--absorption relationship by correlating the emission Stokes parameters \((Q_s, U_s)\) with their optical fractional counterparts \((q_v, u_v)\), which is directly related to $R_{P/p}$ in the idealized case of orthogonal polarization, as shown below. 

The optical Stokes parameters are defined as
$q_v = p \cos 2\psi_v$ and $u_v = p \sin 2\psi_v$, where $\psi_v$ is the polarization angle in
extinction, while $\psi_s$ is used to denote the polarization angle in emission. In the ideal case where
noise is negligible and the polarization pseudo-vectors in extinction and emission are orthogonal
($\psi_s=\psi_v+\pi/2$), one has
\begin{align}
Q_s &= P \cos 2\psi_s
     = -\,P\cos 2\psi_v
     = -\frac{P}{p}\,q_V
     = -R_{P/p}\,q_V, \\
U_s &= P \sin 2\psi_s
     = -\,P\sin 2\psi_v
     = -\frac{P}{p}\,u_V
     = -R_{P/p}\,u_V,
\end{align}
(up to the sign convention adopted for $U$). Thus, the slopes of the correlations $Q_s$ versus
$q_V$ and $U_s$ versus $u_V$ both yield unbiased estimates of the same physical ratio $R_{P/p}$. 
In practice, this correlation analysis is performed on the pixelized maps constructed as explained in Section \ref{sec:pixelization}.

We identify the HEALPix pixels that have at least one stellar polarization measurement. We include the {\textit{Planck}} data at those pixels in our 
analysis.
Following the pixelization procedure outlined above, the resulting pixelized maps for RoboPol and {\textit{Planck}} are shown in Fig.~\ref{fig:gnomonic}. These maps depict the Gnomonic projections of the pixelized Stokes parameters for the sky patch observed by RoboPol and {\textit{Planck}} at \( N_{\text{side}} = 256 \), using stars with distance to the Sun greater than 450~pc.
The maps are centered at coordinates $(l,\,b) = (103.5^\circ,\, 22.5^\circ)$. White pixels indicate regions outside RoboPol's observed sky area. Notably, these maps highlight a clear anti-correlation between the {\textit{Planck}} Stokes parameters and the optical fractional Stokes parameters, with Pearson correlation coefficients 
of the pairs of data $(Q_s,\,q_v)$ and $(U_s,\,u_v)$ being $r_Q = -0.83$ and $r_U = -0.84$, respectively. We explore these correlations in the following sections.


\subsubsection{Fitting Procedure}
\label{sec:fittingproc}

\begin{figure*}[h] 
    \centering
    \includegraphics[width=\textwidth]{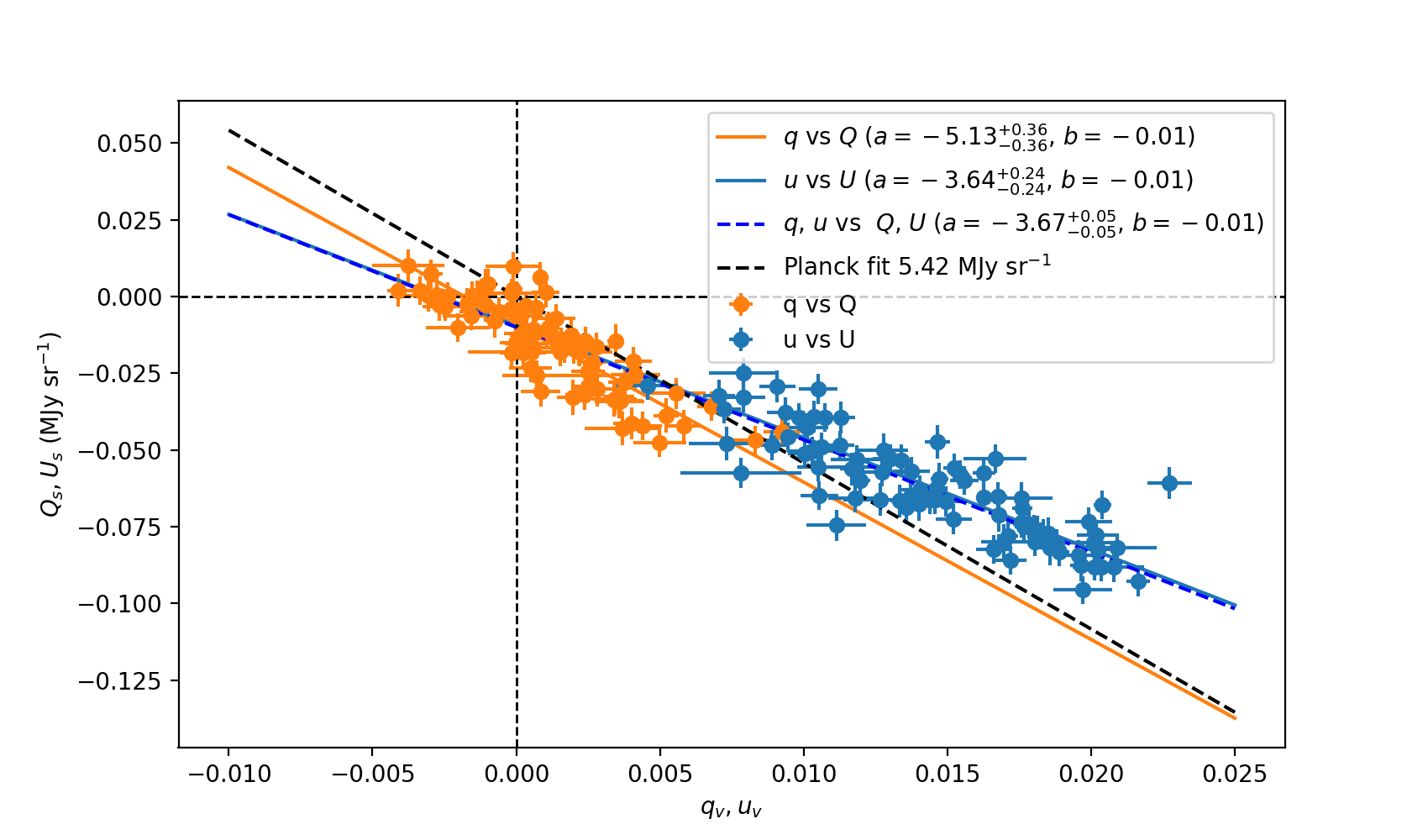}
    \caption{Correlation between the Stokes parameters from {\textit{Planck}} ($Q_S, U_S$) and the fractional Stokes parameters from RoboPol ($q_v, u_v$). Blue points represent the correlation between $U_S$ and $u_v$, while orange points show the correlation between $Q_S$ and $q_v$. The blue dashed line indicates the best-fit line of the joint correlation determined by minimizing $\chi^2$, while solid blue and solid orange correspond to the $U_S-u_v$ and $Q_S-q_v$ fits, respectively. The black dashed line corresponds to the value from \citet{Planck2018}. 
    }
    \label{fig:scatter} 
\end{figure*}

The pixelized polarization data from \textit{Planck} and RoboPol are shown in Fig.~\ref{fig:scatter} in pairs of $(Q_s,\,q_v)$ and $(U_s,\,u_v$) and display a linear relationship, as expected.
%
To study this correlation and determine the ratio of polarized emission to optical polarization $R_{P/p}$, we perform a fit of the form \( y = ax + b \). Here, \( a \) represents the ratio of polarized emission to optical polarization (\( R_{P/p} \)), while \( b \) accounts for any offset between the two datasets.

We perform independent linear fits to the $(Q_s,\,q_v)$ and $(U_s,\,u_v)$ datasets using the Bayesian hierarchical regression method \texttt{LinMix} \citep{Kelly2007}. The prior on the independent variable is represented as a mixture of two Gaussians, \(K=2\), while the slope \( a \)  and intercept \( b \) are assigned broad, non-informative priors.
 Posterior samples of the slope \( a \) and intercept \( b \) are obtained using a Markov Chain Monte Carlo (MCMC) maximum-likelihood analysis. We report the posterior median as the best-fit value, with the posterior standard deviation representing the associated uncertainty.

To estimate a value of \( R_{P/p} \), we also perform a joint fit to both Stokes parameters using a custom MCMC sampler implemented via \texttt{emcee} \citep{ForemanMackey2013}. This approach follows the methodology outlined in \citet{Planck2015stars}. It models the combined system of $(q_v, u_v)$ and $(Q_s, U_s)$ using a shared slope \( a \), a common offset \( b \), and incorporates the full covariance matrix of \textit{Planck} uncertainties ($C_{QQ}$, $C_{UU}$, $C_{QU}$), along with the variances from the stellar measurements. The likelihood is constructed from a chi-square function of the residuals, defined as:

\begin{equation}
\chi^2(a, b) = \mathbf{V}(a, b)^T\mathbf{M}(a, b)^{-1}\mathbf{V}(a, b) \; ,
\end{equation}
where \(\mathbf{V}(a, b)\) and \(\mathbf{M}(a, b)\) are defined as:

\[
\mathbf{V}(a, b) = \begin{pmatrix}
Q_s - aq_v - b \\
U_s - au_v - b
\end{pmatrix},
\]
\[
\mathbf{M}(a, b) = \begin{pmatrix}
C_{QQ} + a^2\sigma_{q_{v}}^2 & C_{QU} + a^2\sigma_{qu_v} \\
C_{QU} + a^2\sigma_{qu_v}  & C_{UU} + a^2\sigma_{u_{v}}^2
\end{pmatrix}.
\]

We report the median and standard deviation of the posterior samples for both the joint fit slope and intercept as our final results.

\section{Results}
\label{sec:results}

\begin{table*}[ht]
    \centering
    \caption{Summary of results from the fitting. 
    The parameters $r_Q$ and $r_U$ are the Pearson correlation coefficients for $Q$ and $U$, respectively.} 
    \label{tab:results_comparison}
    {
    \begin{tabular}{@{}p{2.5cm} c c c c c c c c c@{}}
       \toprule
       & \multicolumn{2}{c}{Best-fit Slope} & Joint fit & & & & & & \\
        Distance cut & \( R_{Qq} \)  & \( R_{Uu} \) & \( R_{P/p}  \) & $b\times 10^{3}$ & $b_q\times 10^{3}$ & $b_u\times 10^{3}$ & $\Delta\Psi_{s/v}$ & $r_Q$ & $r_U$ \\
        (pc) & (${\mathrm{MJy\,sr^{-1}}}$) & (${\mathrm{MJy\,sr^{-1}}}$) & (${\mathrm{MJy\,sr^{-1}}}$) & (${\mathrm{MJy\,sr^{-1}}}$)& (${\mathrm{MJy\,sr^{-1}}}$) & (${\mathrm{MJy\,sr^{-1}}}$) & ($^\circ$) & $-$ & $-$ \\
        \midrule
        \midrule\\[-2.ex]
        > 450 & $ -5.13^{+0.36}_{-0.36}$  
              & $-3.64^{+0.24}_{-0.24}$ 
              & $-3.67^{+0.05}_{-0.05}$ 
              & $-9^{+1}_{-1}$ & $-10^{+4}_{-4}$ & $-10^{+1}_{-1}$ & 4.0 & -0.85 & -0.85 \\
           \addlinespace    
        \addlinespace
        > 2000 & $ -4.09^{+0.51}_{-0.51}$  
               & $-3.11^{+0.30}_{-0.30}$  
               & $-3.83^{+0.08}_{-0.08}$  
                & $-9^{+1}_{-1}$ & $-20^{+5}_{-5}$ & $-9^{+1}_{-1}$ & 3.2 & -0.71 & -0.79 \\
          \addlinespace      
        \addlinespace
        450-2000 & $-5.10 ^{+0.37}_{-0.37}$  
                 & $-3.56^{+0.25}_{-0.25}$  
                 & $-3.63^{+0.06}_{-0.06}$  
                 & $-10^{+1}_{-1}$ & $-11^{+4}_{-4}$ & $-10^{+1}_{-1}$ & 4.1 & -0.83 & -0.83 \\[.5ex]
        \bottomrule
    \end{tabular}
    }
     \label{tab:comparison}
\end{table*}
\subsection{Correlation Analysis }
\label{sec:correlation_analysis}

Figure~\ref{fig:scatter} presents the correlation between the {\textit{Planck}} Stokes parameters (\( Q_s, U_s \)) and the RoboPol fractional Stokes parameters (\( q_v, u_v \)), for stars located beyond 450~pc, as computed in Sect.~\ref{sec:method}.  
We show scatter plots of the paired values, along with best-fit linear relations. Both individual fits to \( Q_s \) versus \( q_v \) and \( U_s \) versus \( u_v \), as well as a combined joint fit to all \((Q_s, U_s)\) versus \((q_v, u_v)\) pairs, are presented. For brevity, we denote the fitted slopes as \( R_{Qq} \), \( R_{Uu} \), and \( R_{P/p}  \), corresponding to the \( Q_s \) vs \( q_v \), \( U_s \) vs \( u_v \), and joint \((Q_s, U_s)\) vs \((q_v, u_v)\) fits, respectively. Table~\ref{tab:comparison} summarizes the results of these fits.

The plots reveal a strong anti-correlation between the datasets. A joint fit yields a slope of \(-3.67 \pm 0.05\) ${\mathrm{MJy\,sr^{-1}}}$, significantly different from the \(-5.42\) ${\mathrm{MJy\,sr^{-1}}}$ value found in \cite{Planck2018}. When fitted separately, the slopes are approximately \(-5.13 \pm 0.4\) ${\mathrm{MJy\,sr^{-1}}}$ for \( Q_s \) versus \( q_v \), and \(-3.64 \pm 0.2\) ${\mathrm{MJy\,sr^{-1}}}$ for \( U_s \) versus \( u_v \). Notably, \( R_{Qq} \) is consistent with the \textit{Planck} result, while \( R_{Uu} \) remains significantly different.

The uncertainty on the slope  \( R_{Qq} \) is higher compared to the joint and \( R_{Uu} \) fits. This can be attributed to two factors: the median signal-to-noise ratio in \( q_v \) is 8 times lower than that of \( u_v \) and the fact that  \( q_v \) data span a narrower range of values than \( u_v \): the interquartile range of \( q_v \) is 0.0028, while that of \( u_v \) is 0.0068. The combined joint fit benefits from the broader range, resulting in a notably smaller uncertainty of 0.05~${\mathrm{MJy\,sr^{-1}}}$r, compared to approximately 0.4 ${\mathrm{MJy\,sr^{-1}}}$ for \( R_{Qq} \). Despite the larger uncertainties, the detected slopes for both \(Q_s\text{--}q_v\) and \(U_s\text{--}u_v\)  relations are statistically highly significant, and the differences between them are not attributable to fitting uncertainties.


In all cases, a small but non-zero intercept is required for an optimal fit. Fixing \( b = 0 \) results in significantly worse fits, as reflected in much higher minimal \(\chi^2\) values. For the \( Q_s \) vs. \( q_v \) relation, the minimum \(\chi^2\) decreases by a factor of 1.61 when allowing \( b \neq 0 \). In the joint fit of \( (Q_s, q_v) \) and \( (U_s, u_v) \), the improvement is much more substantial, with the \(\chi^2\) dropping by a factor of about 35.4. In contrast, for the \( U_s \) vs. \( u_v \) fit, the change is modest (a factor of 1.07), suggesting that the intercept plays a less critical role in that case.

\begin{figure}[h] 
\centering
    \includegraphics[scale = 0.8]{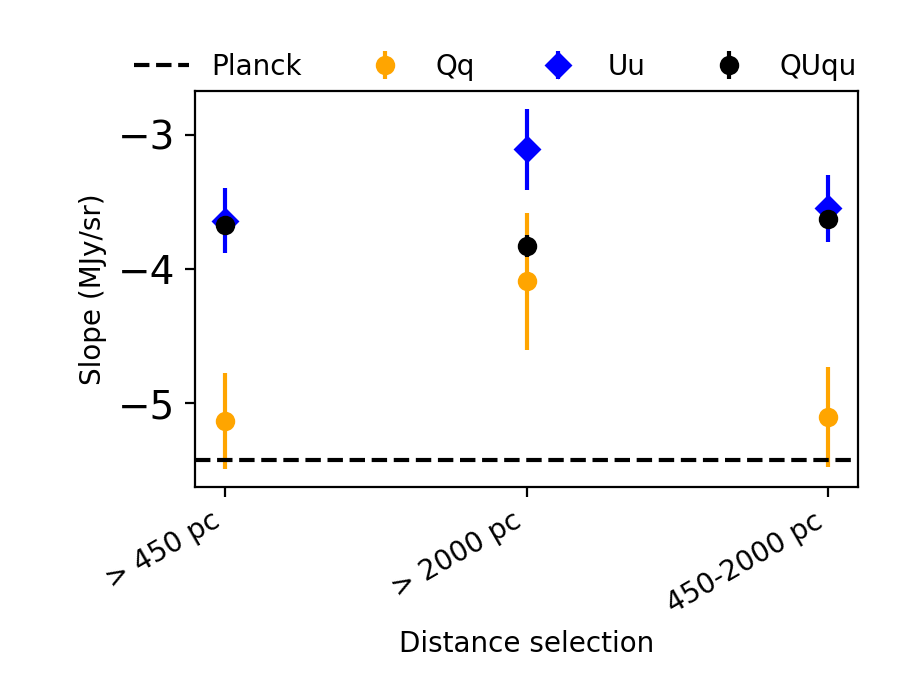} 
    \caption{Comparison of best-fit slope values for different star sample selections. 
    All cases are analyzed at the same $N_{\rm{side}}$ of 256. 
    The slope found in \citet{Planck2018} is shown as the black dashed line.
}
    \label{fig:distance_variation}
\end{figure}

As discussed in Sect.~\ref{sec:method}, 
it is important to check the robustness of our results against the distance-selection criterion. Therefore, we repeat our analysis with different distance selections, the results of which are summarized in Fig.~\ref{fig:distance_variation} where we show
the variation of the fitted slope values as a function of different distance cuts. The slopes \(R_{Uu}\) and \(R_{P/p}\) are mutually consistent within \(1\sigma\) across the distance selections. In contrast, \(R_{Qq}\) varies with the distance cut: the \(>2000\,\mathrm{pc}\) selection yields a slope that is inconsistent with both the \(>450\,\mathrm{pc}\) and \(450\text{--}2000\,\mathrm{pc}\) cuts and with the \textit{Planck} value, whereas the latter two cuts remain mutually consistent (within \(1\sigma\)). These results confirm that our primary conclusion is robust to distance selection: the \(Q_s\text{--}q_v\) and \(U_s\text{--}u_v\) correlations have significantly different slopes, and the joint slope differs from the \textit{Planck} value.




In addition, we repeat the analysis at $N_{\rm{Side}}$ = 512, where we observe a slight flattening of the \( R_{Qq} \) slope compared to the results presented above, while \( R_{Uu} \) and \( R_{P/p} \) remain largely unchanged. We also experimented with different approaches for estimating the mean Stokes parameters per pixel, including the method of \citet{gatz1995weighted} and the inverse-variance weighting described in Sect.~\ref{sec:pixelization_rbpl}. The results of these validation tests were found to be consistent within 1$\sigma$ for the distance cuts of $>450$~pc and $450$--$2000$~pc. However, for the $>2000$~pc cut, we obtain higher values, the details of which are discussed in Appendix~\ref{sec:Nonlinear}.


\subsection{Angle differences with Planck}
\label{sec:angle_diffs}

In the previous section, we noted a difference between the joint slope $R_{P/p}$ in our dataset and that reported by \citet{Planck2015stars} over a larger region of the sky. 
One possible explanation for this discrepancy could be that optical polarization from stars in our sample, may not be probing the entire column of dust traced by the submillimeter polarization from \textit{Planck} along the line of sight (LOS) \citep{Skalidis2019}. 

If the polarized emission arises from the same region that causes the polarization of starlight in absorption, the polarization angles from both types of data should differ by $90^\circ$. 
To test this, we compute the angle differences between starlight and \textit{Planck} polarization for our different distance-cut criteria that select the stars that contribute toward the pixelized RoboPol data.
We estimate the angle difference $\Delta\Psi_{s/v}$ defined from the Stokes parameters and introduced in \citet{Planck2015stars} as\footnote{We note a minus sign difference between our definition and the original one. This comes from our choice of convention for the polarization angle.}
\begin{equation} 
\Delta\Psi_{s/v} = \frac{1}{2} \text{atan2} [(U_s q_v - Q_s u_v), -(Q_s q_v + U_s u_v)] \; .
\label{eqref:DeltaPsi}
\end{equation}
This angle difference is defined in the range -90$^\circ$ to 90$^\circ$ and takes a value of 0 when the polarized emission and absorption are perpendicular.



\begin{figure*}[h!]
    \centering
    \includegraphics[scale = 0.7]{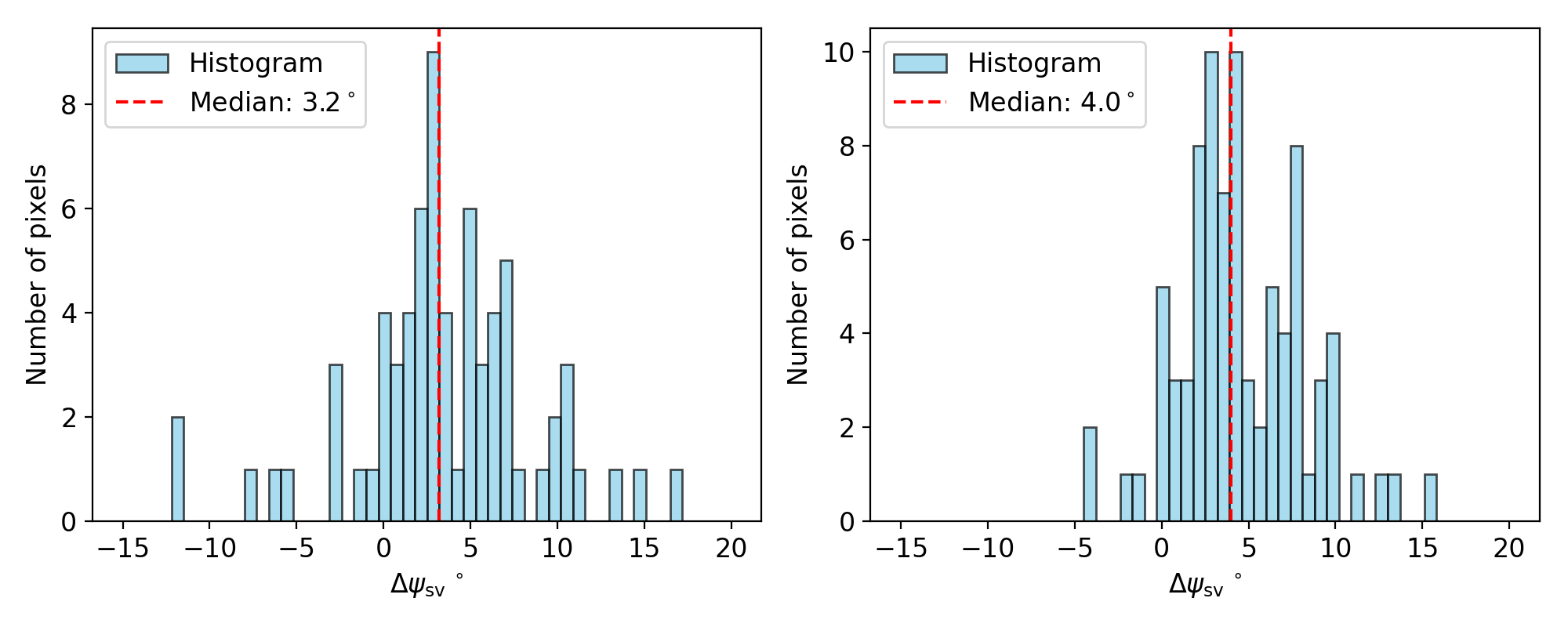}
    \caption{
    Distributions of $\Delta \Psi_{\text{s/v}}$ (in degree) for stars with distance $d \geq 2000$~pc (left) and $d \geq 450$~pc (right). The red vertical lines indicate the distribution's median with values reported in the legend.
    }
    \label{fig:disttribution}
\end{figure*}

We compare the stellar polarization angles, using all the stars located beyond 450\,pc, with those measured by {\textit{Planck}} (after conversion to the IAU convention). The distribution of the angle differences ($\Delta\Psi_{\mathrm{s/v}}$) is shown in Fig.~\ref{fig:disttribution} (right). We find a median offset of 4.0$^\circ$. 
A similar offset between stellar and {\textit{Planck}} polarization angles was found by  \cite{Planck2018}. Their distribution of $\Delta\Psi_{s/v}$ peaked at -3.1$^\circ$, a sign flip relative to our result due to the convention used,  indicating a small systematic shift between the polarization angles measured in emission versus extinction. \cite{Planck2018} commented on the fact that such an offset may arise from differences in angular resolution and probed distances (beam effects) or potential residual systematics in the determination of the zero-point polarization angle of the optical data, or variations of magnetic field/dust properties along the line of sight.

To investigate whether this offset is due to some stars in our sample not tracing the full column, we examine the angle differences for stars located at greater distances (beyond 2\,kpc) compared to the {\textit{Planck}} data. As shown in Fig.~\ref{fig:disttribution} (left), the offset decreases to a median value of 3.2$^\circ$. Accounting for a sign flip, the offset that we observed is small and consistent in magnitude with that reported by \cite{Planck2018}. 
A Kolmogorov–Smirnov (K-S) test between the distributions of $\Delta\Psi_{s/v}$ for the 450-2000 pc versus the $>2000$ pc samples shows that the distributions are consistent with each other (K-S statistic of 0.13 and a $p$-value of 0.44). 
This suggests that the choice of distance cut does not strongly affect the overall $\Delta \Psi_{s/v}$ distribution and that differences in distance alone do not explain the observed angle offset. Furthermore, since stars beyond 2 kpc may lie behind the furthest clouds in the region \citep{Pelgrims2024}, it is unlikely that incomplete column tracing is responsible for the slope discrepancy seen between our fits and those of previous works.

Therefore, the small median angle offset is more likely due to beam mismatch, bandpass  mismatch or residual miscalibration errors, rather than the distance selection of the stars. 
However, we expect beam-related effects to be subdominant at the coarse angular resolution adopted in this work (\texttt{Nside = 256}, $\sim14$~arcmin). We verified this by performing Gaussian-smoothing tests, which did not affect our main results. We also considered the potential impact of bandpass mismatch. Throughout this work, we used the Planck PR4 polarization maps, in which bandpass mismatch is explicitly modeled and corrected during mapmaking. As an additional check, we repeated the analysis using PR3 maps, for which such corrections are less complete, and found consistent results. In particular, the observed difference between $R_{Q/q}$ and $R_{U/u}$ remains unchanged. This leaves residual miscalibration as a more plausible contributor to the observed offset. In addition, variations in dust properties (e.g. composition or alignment efficiency) across clouds may also contribute to its origin.   In the following section, we explore these
possibilities in more detail.

\subsection{Exploring the Origin of the Difference Between $R_{Qq}$ and $R_{Uu}$}

In Sect.~\ref{sec:correlation_analysis}, we found a significant difference between the slopes for the \(Q_s\text{--}q_v\) and \(U_s\text{--}u_v\) and non-zero intercepts, as well as a significantly different $R_{P/p}$ compared to that found by \citet{Planck2015stars,Planck2018}. We have verified that our methodology reproduces results compatible with the \textit{Planck} analysis when applied to that same dataset using two different versions of the \textit{Planck} data (see Appendix~\ref{sec:planck_reproduction}). This confirms that the differences in our findings are not due to methodological choices or changes in the systematic uncertainties of the \textit{Planck} data. The `atypical' $R_{P/p}$ value also cannot arise from the wavelength-dependent variations of the starlight polarization \citep{Serkowski1975}, as the differences between the bandpass of the RoboPol data (R-band) compared to those used in the stellar surveys employed in the analysis of \citet{Planck2015stars,Planck2018} are small.

We propose and explore two possible reasons to explain the differences in slopes---both between our results and those from \textit{Planck}, and between our \(Q_s\text{--}q_v\) and \(U_s\text{--}u_v\) relations, namely a miscalibration of the zero angle and line-of-sight variations of the dust properties.
 

\subsubsection{Effect of Zero-Point Angle Miscalibration}
\label{sec:zero_point_angle}
The method for calibrating the zero-point of the polarization angle differs significantly between the {\textit{Planck}} and optical polarization data. The latter are calibrated against `known' standard stars, and may have residual systematic uncertainties at the level of $1^\circ$ \citep[but can be higher depending on the calibrator stars used,][]{Blinov2023}.

We investigate the potential effect of a miscalibration of the zero-point polarization angle on the correlation between starlight polarization (\( q_v, u_v \)) and dust emission (\( Q_s, U_s \)) using a toy model. We  introduce an offset angle (\( \beta \)) in the polarization plane, and produce mock observations of stellar polarization and polarized dust emission. We then demonstrate that the measured Stokes parameters undergo a systematic rotation, altering the observed slopes and intercepts in the regression analysis.

We generated 500 random realizations of position angles and polarization degrees, drawn from normal distributions: 

\begin{equation}
    \Psi_v \sim \mathcal{N}(\Psi_{B\perp}, \sigma_{\Psi_{B\perp}}), \quad p_v \sim \mathcal{N}(p, \sigma_p).
\end{equation}
For these simulations, we adopted two values of the mean polarization angle: \(\Psi_{B\perp} = 30^\circ\) or \(60^\circ\), with a standard deviation of \(\sigma_{\Psi_{B\perp}} = 5^\circ\). The polarization degree \(p\) was drawn from a normal distribution, with two cases considered: one with a mean of \(2.5\%\) and another with a mean of \(1.5\%\), both with a standard deviation of \(\sigma_p = 0.2\%\). Using these values, we computed the stellar reduced Stokes parameters in the IAU convention as:

\begin{equation}
    \begin{pmatrix} q_v \\ u_v \end{pmatrix} = p_v
    \begin{pmatrix} \cos{[2\Psi_v]} \\ \sin{[2\Psi_v]} \end{pmatrix}.
\end{equation}

Next, we introduce a single $R_{P/p}$ value  (5.42 ${\mathrm{MJy\,sr^{-1}}}$) and compute the corresponding Stokes parameters for dust emission:

\begin{equation}
    \begin{pmatrix} Q_s \\ U_s \end{pmatrix} = R_{P/p} 
    \begin{pmatrix} q_v \\ u_v \end{pmatrix}.
\end{equation}

\begin{figure*}[h] 
   \hspace{-0.5cm} 
\includegraphics[scale = 0.60]{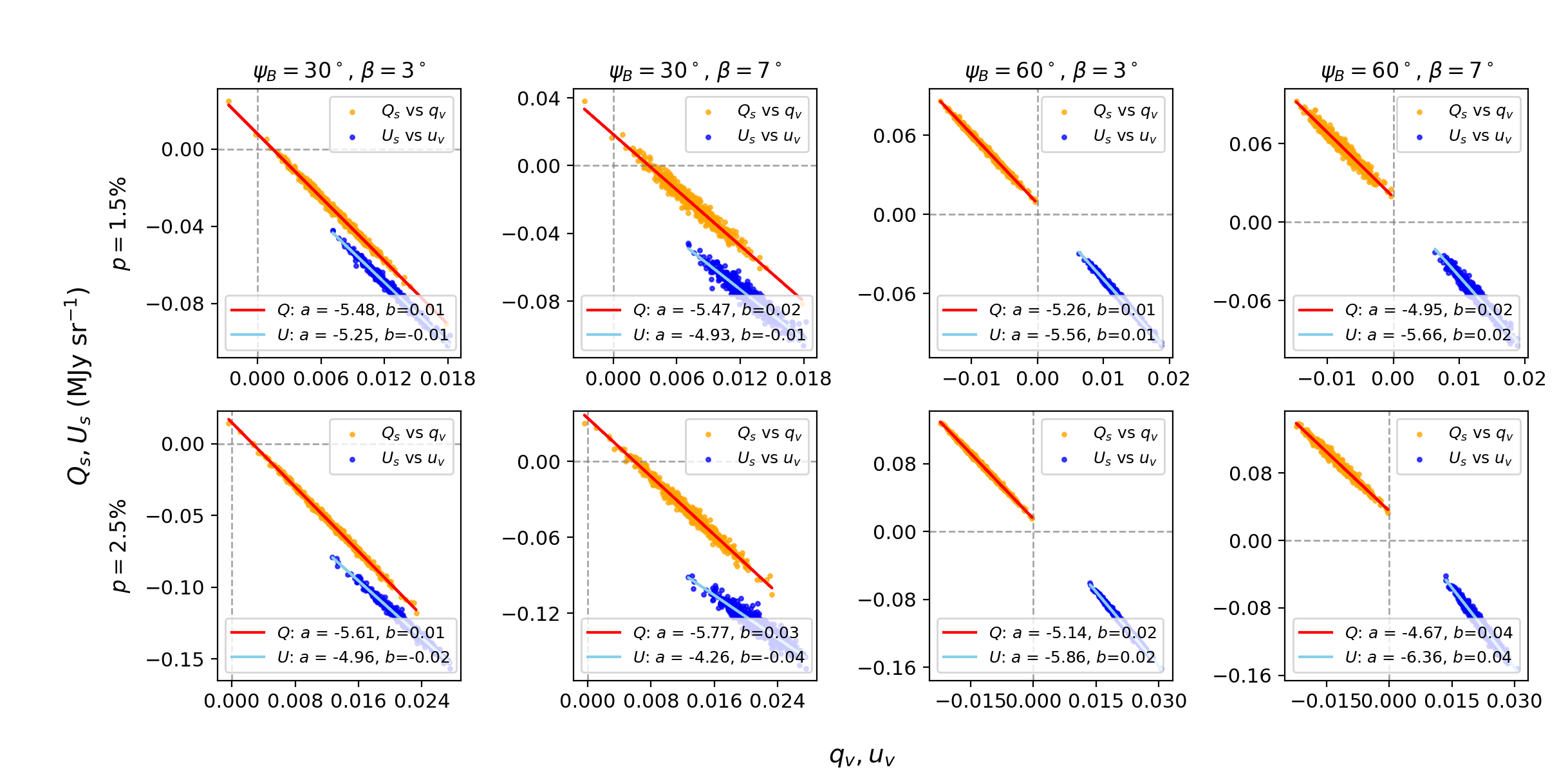}
    \caption{Correlation plots and linear fits between \( (Q_s, q_v) \) and \( (U_s, u_v) \) for two values of the polarization degree, \( p = 1.5\% \) (top row) and \( p = 2.5\% \) (bottom row). Each column corresponds to a combination of magnetic field orientation \( \Psi_{B\perp} = \{30^\circ, 60^\circ\} \) and miscalibration offset \( \beta = \{3^\circ, 7^\circ\} \), increasing from left to right.
 The best-fit values for the slopes (absolute value) and for the intercept obtained for the correlations are given in the legend and the resulting model shown on the scatter plot, in blue for $(Q_s, q_v)$, and orange for $(U_s, u_v)$. They are expressed in MJy sr$^{-1}$. To generate the data, we used $R_p/p = 5.42$ MJy sr$^{-1}$ and $\sigma_{\Psi_{B\perp}} = 5^\circ$, and $\sigma_p = 0.2\%$ to produce some dynamical range.}
    \label{fig:miscalibration1}
\end{figure*}

Then a miscalibration in the zero-point of the polarization angle is introduced. To model this calibration error, we assume that the polarization angle is offset by an angle $\beta$. This results in a rotation of the measured Stokes parameters relative to their true values:

\begin{equation}
    \begin{pmatrix} Q_s \\ U_s \end{pmatrix}_{\text{measured}} =
    \begin{pmatrix} \cos 2\beta & -\sin 2\beta \\ \sin 2\beta & \cos 2\beta \end{pmatrix}
    \begin{pmatrix} Q_s \\ U_s \end{pmatrix}_{\text{true}}.
\end{equation}
We analyze the simulated data in the same way as the actual data, and Fig.~\ref{fig:miscalibration1} illustrates the results, showing both the fitted slopes and the resulting intercepts for various misalignment scenarios. For different values of the magnetic field position angle (\( \Psi_{B\perp} \)), degree of polarization (\( p \)), and miscalibration offset (\( \beta \)), we found that small calibration errors can impact the inferred relationships between \( (Q_s, q_v) \) and \( (U_s, u_v) \). The effect becomes more pronounced for larger miscalibration angles and higher degrees of polarization, leading to variations in the slopes and intercepts of the fitted correlations.
For simplicity, the miscalibration angle that we introduced in our toy model affects only the simulated dust emission data. A similar approach could be applied to the stellar polarization or both. Furthermore, no noise was added to the data in the above exercise. Therefore, when computing the difference in the position angle of the magnetic field inferred from dust emission and from the stars, we always obtain a consistent and unique value equal to the imposed offset. Although the true relation between our simulated sub-mm and optical Stokes parameters is strictly linear and passes through the origin, our fits yield a non-zero intercept. 
These simulations demonstrate that differences in the slopes of the \(Q_s\)--\(q_v\) and \(U_s\)--\(u_v\) correlations can arise from a miscalibration in the zero-point of the polarization angle, particularly when the polarization degree \(p\) is relatively high (e.g., 2.5\%) and the miscalibration angle \(\beta\) is large (e.g., \(7^\circ\); see the fourth column of the second row in  Fig.~\ref{fig:miscalibration1}).
However, in our dataset, the typical polarization degrees are closer to 1.5\% and the miscalbration angle \(\beta\) can be up to \(\sim 3^\circ\) (see Fig.~\ref{fig:disttribution}). This would correspond to the first and third column of the first row in Fig.~\ref{fig:miscalibration1}. Large systematic miscalibration angles are unlikely given current calibration standards \citep{Blinov2023}. 
Therefore, while our model demonstrates that miscalibration can induce an asymmetry between the two correlations, the size of the miscalibration required to reproduce the observed differences in our data appears unrealistically large. This suggests that zero-point angle miscalibration alone is unlikely to fully account for the observed slope differences, and that additional physical or observational factors must be at play.

\subsubsection{Evidence for cloud to cloud variations in $R_{P/p}$}
\label{sec: Multi-cloud Origin}
 Here we present arguments to explain the origin of the difference in slopes between the \(Q_s\)--\(q_v\) and \(U_s\)--\(u_v\) data in terms of a variable $R_{P/p}$ in clouds along the line of sight.

Analyses of stellar polarimetry, distances and H${\rm{~I}}$ emission spectra showed the presence of multiple clouds at different distances in the region under investigation \citep{Panopoulou2019ApJ...872...56P,Pelgrims2023,Pelgrims2024}. In particular, \citet{Pelgrims2024} 
found that the majority of the area is occupied by two clouds at different distances, that overlap on the plane of the sky, producing distinct signatures in the Stokes parameters of the stellar measurements. From their result, we can infer that the observed polarized dust emission (for a given pixel) will be the result of the contribution of the individual cloud contributions:
\begin{align}
Q_{s} &= Q^{(1)} + Q^{(2)}, \\
U_{s} &= U^{(1)} + U^{(2)},
\end{align}
where $Q_{s}, U_{s}$ are the observed Stokes parameters from {\textit{Planck}} and $Q^{(1)}, U^{(1)}$, $Q^{(2)}, U^{(2)}$ are 
their individual contributions from the two clouds having different distances. In what follows, we label the nearest cloud with the index 1, and the farther cloud with index 2.

We also expect that the Stokes parameters of the emission of each cloud will be related to those of the absorption arising from each cloud via a ratio $R_{P/p}$:
\begin{align}
Q^{(1)} &= - R_{P/p}^{(1)}\, q_v^{(1)}, & U^{(1)} &= - R_{P/p}^{(1)}\, u_v^{(1)},\\
Q^{(2)} &= - R_{P/p}^{(2)}\, q_v^{(2)}, & U^{(2)} &= - R_{P/p}^{(2)}\, u_v^{(2)}.
\end{align}
Here, \(q_v^{(1)}\) and \(u_v^{(1)}\) are the stellar Stokes parameters for cloud~1 (with \(Q^{(1)}\), \(U^{(1)}\) the corresponding emission), and \(q_v^{(2)}\) and \(u_v^{(2)}\) are for cloud~2 (with \(Q^{(2)}\), \(U^{(2)}\) the corresponding emission). \(R_{P/p}^{(1)}\) is the ratio for the first cloud and \(R_{P/p}^{(2)}\) the ratio for the second cloud along the line of sight (LOS). We note that dust properties may vary from cloud to cloud; therefore we allow different slopes when modeling different clouds. However, within a given cloud the dust properties are expected to be uniform, and we thus adopt the same slope for both Stokes parameters.


Combining the above sets of equations, we can write the observed Stokes parameters of the emission as:
\begin{align}
\nonumber
Q_{s} &= - R_{P/p}^{(1)} q_v^{(1)} - R_{P/p}^{(2)} q_v^{(2)}\\
U_{s} &= - R_{P/p}^{(1)} u_v^{(1)} - R_{P/p}^{(2)} u_v^{(2)},
\label{eqn:Q_obs}
\end{align}
Up to this point we have made no assumptions about the relationship between the ratios $R_{P/p}^{(1)}$ and $R_{P/p}^{(2)}$ of the two clouds. 

Next, we assume that $R_{P/p}^{(1)} = R_{P/p}^{(2)} = R$ and will show that this leads to a contradiction with the data.
Under this assumption, we can re-write Eq.~(\ref{eqn:Q_obs}) as:
\begin{align}
Q_{s} &= - R (q_v^{(1)} + q_v^{(2)}) = - R q_{obs} \nonumber \\
U_{s} &= - R (u_v^{(1)} + u_v^{(2)}) = - R u_{obs},
\end{align}
Thus, our assumption of the two clouds having the same ratio $R$ implies that we should measure the same slope for the \(Q_s\)--\(q_v\) and \(U_s\)--\(u_v\) data.
This is incompatible with the results obtained in Sect.~\ref{sec:correlation_analysis}, what leads us to conclude that the polarization ratios in cloud 1 and 2 are different ($R_{P/p}^{(1)} \neq R_{P/p}^{(2)}$).





If a tomographic decomposition of the stellar polarization Stokes parameters is available, one can use it to fit for the unknown ratios $R_{P/p}^{(1)}, R_{P/p}^{(2)}$ via Eq.~\eqref{eqn:Q_obs}. We investigate how such information can be used to explain the observed discrepancy between the \(Q_s\text{--}q_v\) and \(U_s\text{--}u_v\) slopes found in Sect.~\ref{sec:correlation_analysis}. 

The tomographic decomposition of \citet{Pelgrims2024} showed that for all pixels in the map:
\begin{equation}
    u_v^{(1)} \ll u_v^{(2)}
\end{equation}
The second cloud appears to dominate the signal in $u_{v}$, due to its particular magnetic field geometry (forming an angle of $\sim 40^\circ$) and its higher $p$ compared to the first cloud (1.4\% compared to 0.2\%).
This allows us to simplify the system in Eq.~\eqref{eqn:Q_obs}. If we assume that $R_{P/p}^{(1)}$ and $R_{P/p}^{(2)}$ are of the same order of magnitude (so that $R_{P/p}^{(1)}  u_v^{(1)} \ll R_{P/p}^{(2)}  u_v^{(2)}$) we obtain:
\begin{align}
Q_{s} &= - R_{P/p}^{(1)} q_v^{(1)} - R_{P/p}^{(2)} q_v^{(2)}\\
U_{s} &\approx - R_{P/p}^{(2)} u_v^{(2)},
\end{align}
Because in this region $u_v \approx u_v^{(2)}$, the second of these equations shows that the observed \(U_s\text{--}u_v\)  slope in our dataset is representative of the second cloud.


We verify this conclusion by creating mock data 
for our two-cloud hypothesis, where the clouds may possess different intrinsic properties—such as porosities, temperature or shape \citep{Draine2021} — that result in distinct $R_{P/p}$ values and, consequently, affect the observed correlations. The observed Stokes parameters are expressed as the sum of contributions from two separate clouds:

\begin{equation}
    \begin{pmatrix} q_v \\ u_v \end{pmatrix} = \begin{pmatrix} q_v^{(1)} \\ u_v^{(1)} \end{pmatrix} + \begin{pmatrix} q_v^{(2)} \\ u_v^{(2)} \end{pmatrix}
\end{equation}

\begin{equation}
    \begin{pmatrix} Q_s \\ U_s \end{pmatrix} =  - R_{P/p}^{(1)} \begin{pmatrix} q_v^{(1)} \\ u_v^{(1)} \end{pmatrix} - R_{P/p}^{(2)} \begin{pmatrix} q_v^{(2)} \\ u_v^{(2)} \end{pmatrix}
\end{equation}

The relative contribution from each cloud depends on the degree of polarization ($p_v^{(i)}$) and the plane-of-sky magnetic field orientations in both absorption and emission. 
We chose the values for the degree of polarization ($p^{[1]}$,$p^{[2]}$) and polarization angles ($\Psi_{B\perp}^{[1]}$,  
$\Psi_{B\perp}^{[2]}$) based on the tomographic decomposition presented by \citet{Pelgrims2024}. We adopt an $R_{P/p}$ value for the second cloud that is equal to the slope found in the $U_s-u_v$ correlation. For the first cloud, we adopt $R_{P/p} = 5.4$~${\mathrm{MJy\,sr^{-1}}}$, similar to the value found by {\textit{Planck}}. 

\begin{figure}[h] 
   \centering
\includegraphics[scale = 0.5]{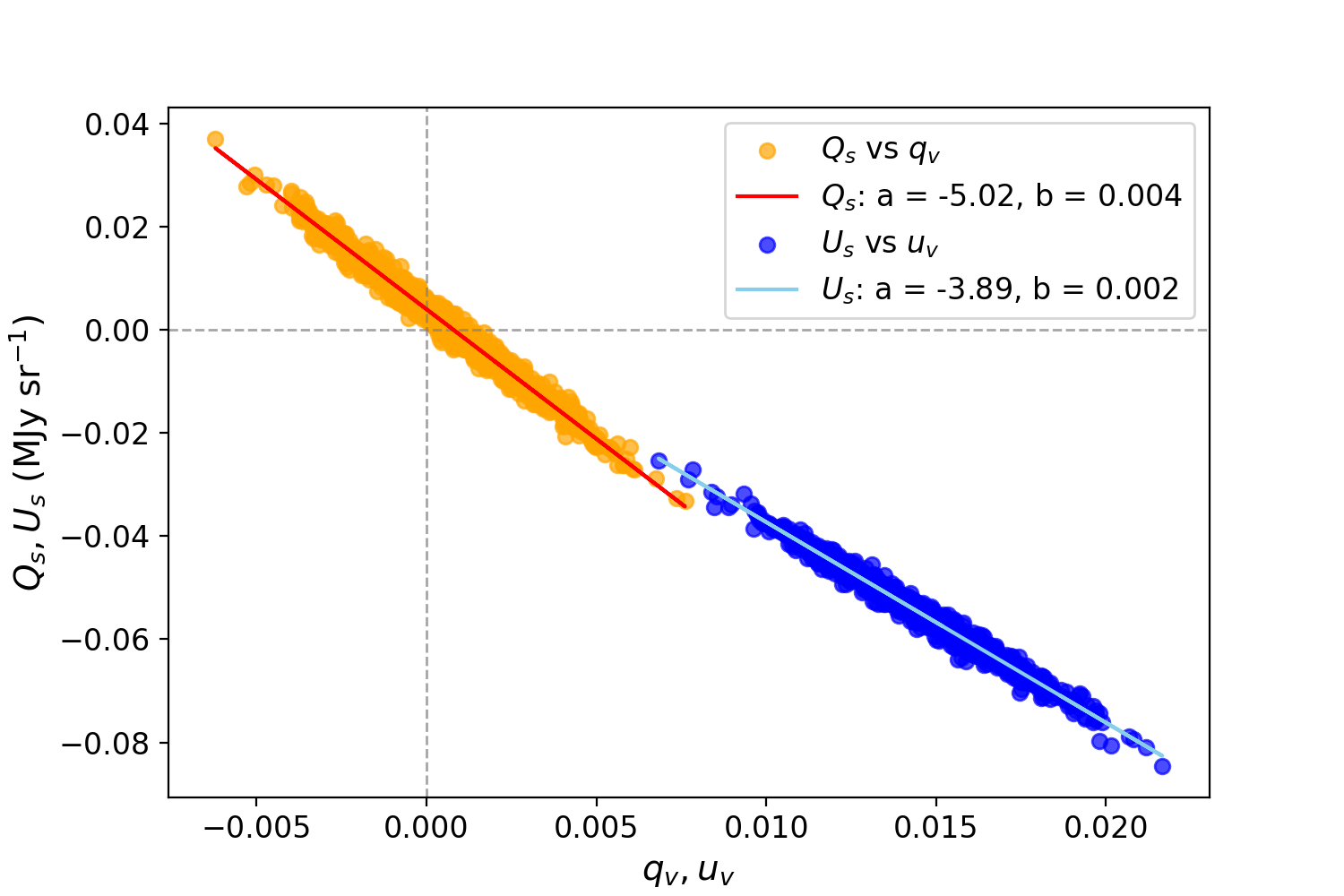}
   
    \caption{Correlation plot for the simulated example of two intervening clouds with different polarization
properties and different $R_{P/p}$ ratios, as described in the text.
}
    \label{fig: multicloudcorrelation}
\end{figure}

Fig.~\ref{fig: multicloudcorrelation} illustrates an example of the resulting correlation plot for a case with  
$p^{[1]} = 0.2\%$, $p^{[2]} = 1.4\%$, $\Psi_{B\perp}^{[1]} = 80^\circ$,  
$\Psi_{B\perp}^{[2]} = 40^\circ$, and $R_{P/p}^{[1]} =  5.4$, $R_{P/p}^{[2]} =  3.7$. We generate 500 synthetic measurements (without added noise) using  
$\sigma_{\Psi_{B\perp}}^{[i]} = 2^\circ$, $\sigma_p^{[1]} = 0.2\%$, and  
$\sigma_p^{[2]} = 0.2\%$. 
This example produces results qualitatively similar to the data of Fig.~\ref{fig:scatter}. 
The measured slope for $(U_s, u_v)$ is close to the value assumed for cloud 2.  

A non-zero intercept also arises in this case, as illustrated in Figure \ref{fig: multicloudcorrelation}. This becomes evident when we express Eq.~\ref{eqn:Q_obs} in terms of the observed Stokes parameters. For example:
\begin{align}
Q_{s} &= - R_{P/p}^{(1)} q_v^{(1)} - R_{P/p}^{(2)} q_v^{(2)} \\
      &= - R_{P/p}^{(1)} q_v^{(1)} - R_{P/p}^{(2)} (q_{\text{obs}} - q_v^{(1)}) \\
      &= - \left( R_{P/p}^{(1)} - R_{P/p}^{(2)} \right) q_v^{(1)} - R_{P/p}^{(2)} q_{\text{obs}}.
\end{align}

Here, \( q_{\text{obs}} \) is the total observed stellar Stokes parameter. When expressing \( Q_s \) as a function of \( q_{\text{obs}} \), the mismatch between the relative contributions of the two clouds introduces an intercept term proportional to \( \left( R_{P/p}^{(1)} - R_{P/p}^{(2)} \right) q_v^{(1)} \), even though the underlying physical relation among the components contains no intercept.

This analysis confirms that variations in $R_{P/p}$ across multiple clouds can produce different $R_{Qq}$ vs $R_{Uu}$. The  variations in slopes and intercepts combined with a tomographic decomposition of the optical polarization Stokes parameters could be used to infer properties of the dust, grain alignment and magnetic fields as they vary along the line of sight.

\section{Discussion}
\label{discussion}


In Sect.~\ref{sec:results}, we investigated the correlation between optical and submillimeter polarization Stokes parameters in a localized region of the diffuse ISM. Interestingly, not only did we find a value that differs from that reported previously in different sky regions \citep{Planck2015stars,Planck2018}, but we also observed different slopes for the $R_{Qq}$ and $R_{Uu}$ correlations. We explored various potential explanations for this discrepancy, ruling out the possibility that optical polarization does not trace the full column of dust probed by submillimeter polarization along the LOS (Sect.~\ref{sec:angle_diffs}). We suggest a combination of low-level zero-point angle miscalibration (Sect.~\ref{sec:zero_point_angle}) and variations in dust properties along the LOS (Sect.~\ref{sec: Multi-cloud Origin}) as a likely origin of 
the trend reported in Sect.~\ref{sec:correlation_analysis}.

In \citet{Planck2015stars} only a slight difference in the slopes of ($Q_s$, $q_v$) and ($U_s$, $u_v$) was found, along with the possibility of a small intercept. They compared starlight polarization and dust emission along translucent sightlines at low Galactic latitude, finding slopes of $R_{P/p} = (-5.17 \pm 0.09)$ MJy~sr$^{-1}$ and $(-5.07 \pm 0.13)$ MJy~sr$^{-1}$ for the two correlations. The corresponding y-intercepts were found to be $(0.0077 \pm 0.0013)$ MJy~sr$^{-1}$ and $(-0.0016 \pm 0.0013)$ MJy~sr$^{-1}$. These differences are small compared to the associated uncertainties. More significant differences between $R_{Qq}$ and $R_{Uu}$ were found by \cite{versteeg2025} when comparing NIR stellar polarimetry from \citet{clemens2020} with polarized dust emission from \textit{Planck}.


\subsection{On the atypical $R_{P/p}$ }
In Sect.~\ref{sec: Multi-cloud Origin}, we demonstrated using simulated data how clouds with different values of $R_{P/p}$ can effectively reproduce our observational results. Based on these findings, the question arises of why the $R_{P/p}$ value of the second cloud is so much lower than the typical value of 5.4 ${\mathrm{MJy\,sr^{-1}}}$. Indications of a lower $R_{P/p}$ of $4.2 \pm 0.1 \, {\mathrm{MJy\,sr^{-1}}}$ in the diffuse ISM were also found in a different region of the sky \citep{Panopoulou2019}. 


First, we expect that the observation of a lower value of $R_{P/p}$ (in our study $\sim 3.7$ ${\mathrm{MJy\,sr^{-1}}}$) compared to the typical value, is likely due to an increased optical polarization relative to FIR polarized intensity. The polarized intensity spectrum depends mostly on the dust temperature $T_d$. At 353 GHz the polarized intensity is proportional to the grain temperature: $P \propto T_d$. The grain temperature relates to the strength of the radiation field, $J$, as  $J \propto T_d ^{5.5}$ \citep{Draine2011piim.book.....D}. In order for $R_{P/p}$ to decrease by a factor of 1.5, the radiation field would have to decrease by a factor of almost 10, which seems unrealistic for the diffuse ISM. 

Another factor affecting both the optical and FIR polarization is the minimum size of aligned grains, which determines the mean alignment efficiency \citep[e.g.][]{Hensley2023}. In the FIR, the amplitude of the polarized intensity spectrum increases proportionally to the mean alignment efficiency. The shape of the optical polarization spectrum is also sensitive to the mean alignment efficiency, in a more complex way \citep[e.g.][]{Reissl2020}. We have investigated the effect of changing the minimum size of aligned grains on $R_{P/p}$ in the context of the Astrodust model \citep{Hensley2023}. We find that varying the minimum size of aligned grains in the range of 0.01 - 0.08 $\mu$m has a minimal effect, far too small to explain the observed change in $R_{P/p}$ of a factor of 1.5. 

The shape of the optical polarization curve is sensitive to the shapes and porosities of grains while the polarized intensity spectrum is largely insensitive to these variations \citep[e.g.][]{Voshchinnikov2012rev, Guillet2018,Draine2021}. 
Given this, one might wonder whether there is independent evidence that dust properties (such as their size distribution) are `atypical' in the region under investigation. Indeed, the cloud with anomalous $R_{P/p}$ is part of the Cepheus Flare where the ratio of total to selective extinction, $R_V$, appears to be lower than the mean diffuse ISM value of 3.1: based on the map of \citet{zhang2025} we find a mean $R_V$ in the region at 400-500 pc of 2.6. Since $R_V$ is known to reflect the grain size distribution \citep{MRN1977,Mathis1981,Kim1994}, the lower $R_V$ could indicate an increased population of smaller grains. 

However, in the translucent ISM with $1 \leq A_V \leq  3\, \rm mag$, variations in \( R_V \) may instead be driven by changes in the abundance of Polycyclic Aromatic Hydrocarbons \citep[PAHs;][]{zhang2025pah}, which affect extinction but not polarization. The cloud under investigation is even more diffuse, with extinction ranging from 0.4 - 0.5 mag between 400 and 1000 pc based on the \citet{zhang2025} map. Changes in the PAH population would not reflect in $R_{P/p}$, which mainly traces the aligned, larger grains \citep{Hensley2023}. 

Thus, the low $R_{P/p}$ value in our study suggests the presence of a population of aligned dust grains in this region that are atypical compared to those in the general ISM. At the same time, it is unclear whether the extinction curve supports such a different population of grains. To test our hypothesis, multi-wavelength observations of the optical polarization of stars tracing the cloud would be invaluable. By measuring the wavelength-dependence of the optical polarization curve, one would be able to estimate the wavelength where maximum polarization occurs, $\lambda_{max}$. Correlations between $\lambda_{max}$ and \( R_V \) \citep{whitet} have been linked to variations in the sizes and properties of aligned grains. If the linear correlation between \( R_V \) and $\lambda_{max}$ found by \citet{Serkowski1975} and \citet{whitet} holds\footnote{This may not be the case within clouds \citep[][]{Whittet2001_taurus, Andersson2007, Bartlett2025arXiv250904427B}.} and if it can be extrapolated to lower $A_V$ (these studies typically used stars at $A_V \sim 1$ mag or higher), then an $R_V$ of 2.6 would imply a $\lambda_{max}$ of 0.46 $\mu$m in this cloud, compared to the typical 0.55 $\mu$m value. To properly estimate the $\lambda_{max}$ of the dominant cloud along the LOS, one would also have to correct for the effect of the foreground cloud on the wavelength-dependence of the polarization \citep{Mandarakas2025A&A...698A.168M,Skalidis2024arXiv241108971S}. To explore a potential connection between \( R_V \) and \( R_{P/p} \) multi-band optical polarization measurements are required across many more regions.

\subsection{Assessing the Impact of Small-Scale Variations}
\label{sec:chisq_method}
In the model for estimating the stellar polarization within a given pixel, we make several assumptions. In particular, we assume that the measurements within each pixel contain no additional sources of variability beyond instrumental noise, and that this noise is correctly estimated by the RoboPol pipeline.

To test this assumption, we perform a goodness-of-fit test before rotating the measurements (so that the $q$--$u$ covariance matrices remain diagonal). Specifically, for each individual pixel,
we compute the reduced chi-squared statistic:
\begin{equation}
\label{eq:chisquared}
\chi^2_{\mathrm{dof}} = \frac{1}{2(N-1)}\sum_{i=1}^{N}(\mathbf{d}_i - \langle\mathbf{d}\rangle)^T \mathbf{C}_{qu, i}^{-1} (\mathbf{d}_i - \langle\mathbf{d}\rangle),
\end{equation}
where $\mathbf{d}_i$ is the individual measurement vector
of the stars, $\langle\mathbf{d}\rangle$ is the pixel-averaged value, and $\mathbf{C}_{qu, i}$ is the noise covariance matrix for each stellar measurement.

If the noise is properly characterized and dominates over any intrinsic scatter, we expect $\chi^2_{\mathrm{dof}} \approx 1$. However, we find
that the median value of $\chi^2_{\mathrm{dof}}$ over the full area to be approximately 2.46. The corresponding value for the
\textit{Planck} data is very close to unity. This suggests that the stellar data, probing pencil-beams, reveal significant intrinsic scatter within pixels, whereas the \textit{Planck} data remains noise-dominated.

To account for this apparent underestimation of the uncertainties in the stellar data, we explore inflating the errors by adding an intrinsic scatter term to the covariance matrix:
\begin{equation}
\mathbf{C}_{qu, i} \rightarrow \begin{bmatrix}
\sigma_{q,i}^2 + v_c^2 & 0 \\
0 & \sigma_{u,i}^2 + v_c^2
\end{bmatrix},
\end{equation}
where $v_c$ is chosen such that Eq.~(\ref{eq:chisquared}) yields $\chi^2_{\mathrm{dof}} \approx 1$.

While this approach helps in principle, it oversimplifies the problem. Intrinsic scatter may not  affect the diagonal terms ($C_{qq}^{\mathrm{intsc}}$, $C_{uu}^{\mathrm{intsc}}$)
equally and can also
introduce a non-zero off-diagonal term ($C_{qu}^{\mathrm{intsc}}$), though typically of smaller amplitude (see Appendix~B of \citealt{Pelgrims2023}  for a detailed toy-model based discussion on this matter). Estimating the full intrinsic covariance matrix for each pixel would require numerically solving for all three components ($C_{qq}^{\mathrm{intsc}}$, $C_{uu}^{\mathrm{intsc}}$, and $C_{qu}^{\mathrm{intsc}}$) by maximizing the log-likelihood for each pixel. If one assumes Gaussian statistics for the intrinsic scatter in the $q$ and $u$ measurements, the problem can be somewhat simplified \citep{Skalidis2018A&A...616A..52S}, however the reliability of this assumption needs further testing. Therefore, due to the complexity of this method and the lack of a robust way to constrain all three elements of the covariance matrix reliably for each pixel, we opted not to implement the full error inflation scheme. 

We did, however, test the simplified inflation approach described above and found that the resulting slopes were consistent with those obtained using the un-inflated uncertainties in Sect.~\ref{sec:correlation_analysis}, with deviations well within \(1\sigma\). This confirms that our main results are robust against modest underestimation of uncertainties. Given the small impact on the final results, we adopt the un-inflated inverse-variance approach as our default method for simplicity. Future improvements to this analysis could incorporate full intrinsic covariance estimation and more realistic modeling of small-scale correlations.



\section{Summary}
\label{summary}
In this paper, we estimated the ratio of polarized thermal dust emission to optical starlight polarization ($R_{P/p}$) using a 
recently published 
sample of stellar polarimetry at 
intermediate Galactic latitudes densely sampling a region of approximately four square degrees. The three dimensional structure of the dusty magnetized ISM was previously studied in this relatively diffuse ISM region \citep{Pelgrims2024}. Here our goal was to measure $R_{P/p}$ in this localized region and compare it with that found in larger sky regions by the \textit{Planck} collaboration.

We combined submillimeter polarization data from the \textit{Planck} satellite ({\textit{Planck}} Collaboration, PR4) with optical starlight polarization measurements from the RoboPol survey \citep{Pelgrims2024}.
To minimize discrepancies between the dust traced in absorption and emission, we conducted our analysis adopting three distance selection criteria, considering stars located at distances 
$>$450\,pc, $>$2000\,pc, and between 450--2000\,pc.
This guarantees that the stellar polarization captures most of the dust-induced polarization signal as the dominant polarizing clouds in this sky region were found to be at distances smaller than 450~pc \citep{Pelgrims2024}.
We constructed pixelized maps by computing the weighted mean and its associated error, grouping stars (or \texttt{$N_{\rm{side}}=2048$} pixels) that fall within the same \texttt{$N_{\rm{side}}=256$} pixel for the optical and {\textit{Planck}} data, respectively. 

Our results show significant deviations from those reported by previous studies of $R_{P/p}$. First, the slope of the (\(Q_s,U_s\)) vs (\(q_v,u_v\) ) fit (\(-3.67 \pm 0.05\)\, ${\mathrm{MJy\,sr^{-1}}}$) differs notably from the value of $-5.42$ ${\mathrm{MJy\,sr^{-1}}}$ found in \citealt{Planck2015stars,Planck2018}. Second, the \(Q_s\)--\(q_v\) ($-5.13 \pm 0.4$\, ${\mathrm{MJy\,sr^{-1}}}$) slope differs significantly from the \(U_s\)--\(u_v\) ($-3.64 \pm 0.2$\, ${\mathrm{MJy\,sr^{-1}}}$) and joint slopes.

To investigate the origin of these discrepancies, we explored two possible explanations. First, we considered the effect of miscalibration in the zero-point polarization angle. We simulated how the slopes of \(Q_s\)--\(q_v\) and \(U_s\)--\(u_v\) vary with different degrees of polarization and angle offsets ($\beta$).
The values of the calibration error angles that would be needed to explain our observation were large compared with current estimates, making this first scenario unlikely.
Second, we considered the possibility of a variable $R_{P/p}$ along the line of sight, due to multiple dust clouds with
differing dust properties.
We found that, in this region of the sky, one cloud dominates the correlation in the \(U_s\)--\(u_v\) plane, while both clouds contribute to the correlation in the \(Q_s\)--\(q_v\) plane. Tests with mock data showed that this scenario can effectively reproduce the observed results.

We concluded that spatial variations in \(R_{P/p}\) across multiple clouds are likely responsible for the observed variations in the polarization correlations. 
When multiple clouds contribute along the line of sight, the slopes of the \(Q_s\)--\(q_V\) and \(U_s\)--\(u_V\) correlations are not necessarily tied to a single intrinsic \(R_{P/p}\). Instead, these correlation slopes should be treated as free parameters, which may nevertheless inform \(R_{P/p}\) with appropriate modeling.

Thus, studying such variations over larger areas of the sky will be essential for characterizing dust grain properties in the ISM and understanding their environmental dependence. This, in turn, has important implications for improving foreground modeling and component separation in future CMB polarization experiments. 
For example, optical polarization surveys combined with \textit{Gaia} will enable the construction of tomographic maps of starlight polarization with significant sky coverage \citep{2012AIPC.1429..244M,tassis2018pasiphaehighgalacticlatitudehighaccuracyoptopolarimetric}.  The type of modeling presented in this work could then be expanded to infer the intrinsic $Q_s, U_s$ of clouds along the line of sight, extending the success of using 3D dust extinction maps as templates for foreground modeling \citep{Martinez2018MNRAS.476.1310M,2026arXiv260110640S}. Such tomographic maps of polarized dust emissions would allow us to replace 3D models of plausible sky realizations \citep{2025ApJ...991...23P} with models of the actual polarized sky constructed from independent data. These new models would  help overcome some of the limitations of current models (e.g., \citealt{Vacher2025}), at least by reducing and refining the range of possibilities.
Additionally, the advent of next-generation CMB experiments (e.g., \citealt{2019BAAS...51g.147L}; \citealt{2023PTEP.2023d2F01L}; \citealt{2019arXiv190704473A}) will enable extending our analysis to higher resolution and more frequencies \citep{Hensley2022}, making it possible to infer interstellar dust properties in greater detail (e.g., \citealt{Draine2003ARA&A}; \citealt{Hensley2021}).

\begin{acknowledgements}
N.M. acknowledges support from the National Research Foundation (NRF) of South Africa, the South African Astronomical Observatory, and in part by the U.S. National Science Foundation (NSF) under grant AST2109127. G.P. acknowledges support from the Swedish Research Council (VR) under grant number 2023-04038 and the Knut and Alice Wallenberg Foundation Fellowship program under grant number 2023.0080. 
V.P. acknowledges funding from a Marie Curie Action of the European Union (grant agreement No. 101107047).
This work was carried out, in part, at the Jet Propulsion Laboratory, California Institute of Technology, under a contract with the National Aeronautics and Space Administration. This research is funded by the European Union. Views and opinions expressed are, however, those of the author(s) only and do not necessarily reflect those of the European Union or the European Research Council Executive Agency. Neither the European Union nor the granting authority can be held responsible for them. This work is supported by ERC grant mw-atlas project no. 101166905. N.U. acknowledges support from the European Research Council (ERC) under the Horizon ERC Grants 2021 programme under grant agreement No. 101040021. K.T. acknowledges the support by the TITAN ERA Chair project (contract no. 101086741) within the Horizon Europe Framework Program of the European Commission. The authors acknowledge Interstellar Institute's program "II7" and the Paris-Saclay University's Institut Pascal for hosting discussions that nourished the development of ideas behind this work. Support for this work was provided by NASA through the NASA Hubble Fellowship grant \#~HST-HF2-51566.001 awarded by the Space Telescope Science Institute, which is operated by the Association of Universities for Research in Astronomy, Inc., for NASA, under contract NAS5-26555.

\end{acknowledgements}


\bibliographystyle{aa}
\bibliography{main} 

\appendix
\section{Reproduction of previous results}
\label{sec:planck_reproduction}

We tested our methodology of pixelizing the Stokes parameters (Sect.~\ref{sec:method}) on the sample of stars used in the analysis of \citet{Planck2018}. We compared polarized dust emission at 353\,GHz with optical starlight polarization data from \citet{berdyugin2014}. To ensure a representative sample, we selected stars within the latitude range of 30$^\circ$ to 90$^\circ$ and with a signal-to-noise ratio greater than 3. This criterion left us with a sample of 1674 stars. 

\begin{figure}[h] 
\centering
\includegraphics[scale = 0.52]{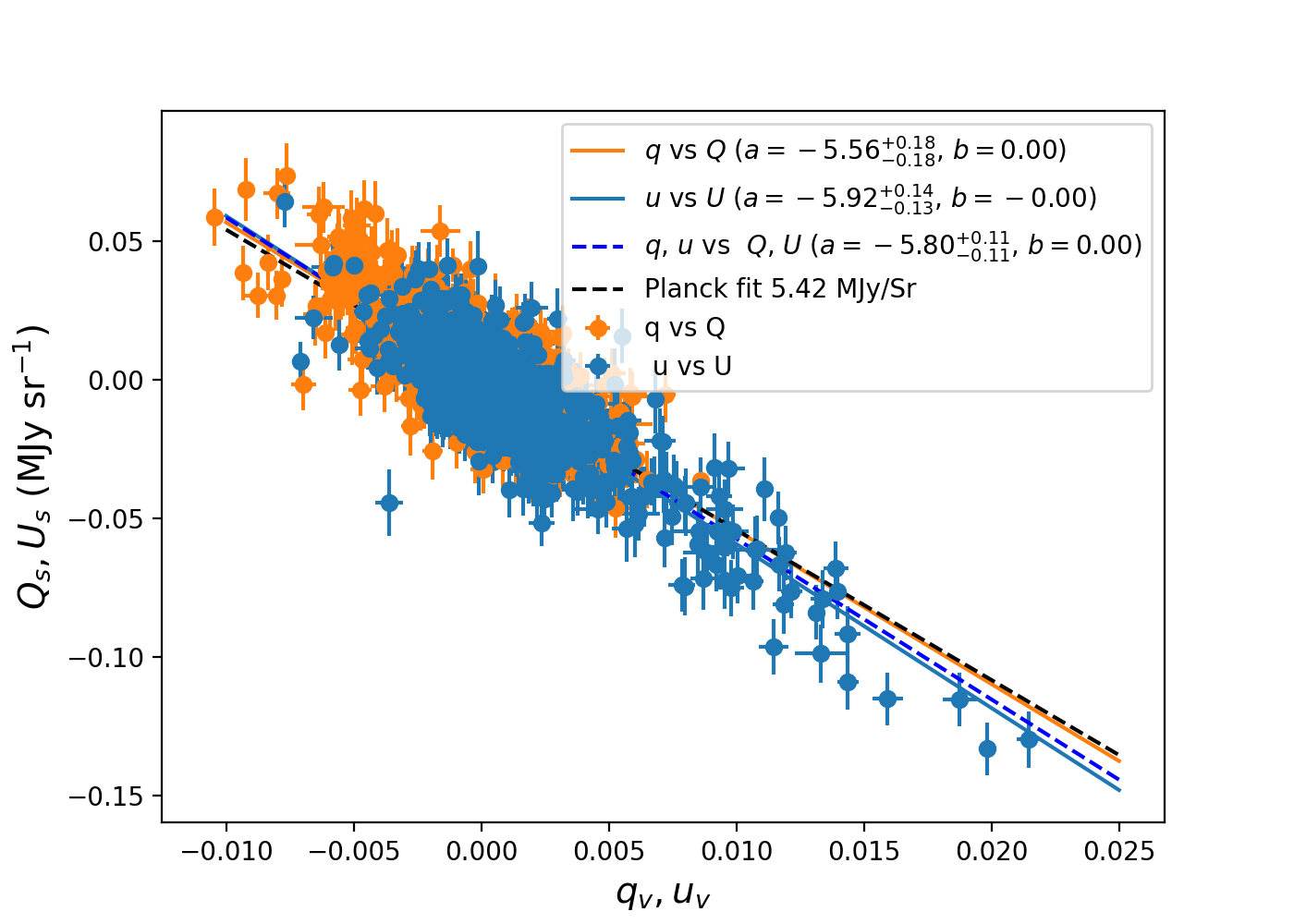}

    \caption{Reproduction of {\textit{Planck}} results with a stellar polarization sample used in \citet{Planck2018} but using the methodology described in Sect.~\ref{sec:method}.
}
    \label{fig:Planck}
\end{figure}

The catalog provides debiased polarization values, which we converted back to the corresponding biased values. 
We then computed the Stokes parameters from the biased degree of polarization and polarization angles, and propagated the associated uncertainties accordingly. For each selected star, the Stokes parameters \(q_v\) and \(u_v\) were computed following the IAU convention. Figure \ref{fig:Planck} illustrates our reproduction of the results of \citet{Planck2018}.

There is a small difference between our derived polarization ratio $R_{P/p} = 5.79$ (dashed blue line in Fig.~\ref{fig:Planck}) and the value reported by the \textit{Planck} collaboration, $5.42$ (dashed black line). We performed the analysis using the Planck PR3 and the Planck PR4 data. We find minor differences in the derived slopes between the two choices of dataset. In particular, the slope of the $Q_s-q_v$ relation changes from -5.59$\pm0.19$ ${\mathrm{MJy\,sr^{-1}}}$ (PR3) to $-5.56_{-0.18}^{+0.18}$ ${\mathrm{MJy\,sr^{-1}}}$ (PR4), that of $U_s - u_v$ changes from $-5.88_{-0.13}^{+0.15}$ ${\mathrm{MJy\,sr^{-1}}}$ (PR3) to $-5.92_{-0.13}^{+0.14}$ ${\mathrm{MJy\,sr^{-1}}}$ (PR4) and the joint fit slope changes from $-5.79\pm0.11$ ${\mathrm{MJy\,sr^{-1}}}$ to $-5.80\pm0.11$ ${\mathrm{MJy\,sr^{-1}}}$ (PR4). Systematic uncertainties alone are too small to explain the difference between our best-fit slope and that found by \citet{Planck2018}.

We attribute this difference to several factors. First, unlike \citet{Planck2018}, we have not corrected the stellar Stokes parameters for the background ISM contribution, that is, we have not derived the $q^{\infty}, u^{\infty}$ values that estimate the polarization assuming the star is located at infinity. Second, it could be due to our line-of-sight selection, as we did not explicitly exclude sightlines with low $E(B - V)_{\star} / E(B - V)_{\infty}$ ratios. Finally, we did not apply a beam depolarization correction to the \textit{Planck} maps. Despite these differences in methodology, the agreement of our fitted relations shown in Fig. \ref{fig:Planck} with the results of \citet{Planck2018} is remarkably good.

\section{ Non-linearity in the $Q_s$--$q_v$ relationship}

\label{sec:Nonlinear}

An important aspect of our analysis is the apparent non-linearity in the correlation between the \textit{Planck} Stokes parameter \( Q_s \) and the corresponding stellar Stokes parameter  \( q_v \). A close inspection of the fit presented in Fig. \ref{fig:scatter} reveals non-linearities in the residuals. 

A striking signature of this non-linearity is seen if we mask the middle portion of \(Q_s - q_v\) datasets in Fig. \ref{fig:scatter}. When excluding the central values, the resulting linear fits no longer pass symmetrically through the data: points with large negative \( q_v \) tend to lie above the fit, while those with large positive \( q_v \) lie below. This behavior violates the basic expectation from a linear model. 


We interpret this non-linearity as a consequence of the varying relative contribution of multiple clouds along the line of sight. When the relative contribution of each cloud changes (e.g., due to a distance cut or sky region selection), the resulting \( Q_s\)--\( q_v \) relationship reflects a mix of different effective slopes. The \( Q_s \) signal depends more heavily on the balance between cloud 1 and cloud 2 contributions than the \( U_s\) signal, which appears more stable across distance cuts.

\begin{figure}[h] 
\centering
\includegraphics[scale = 0.85]{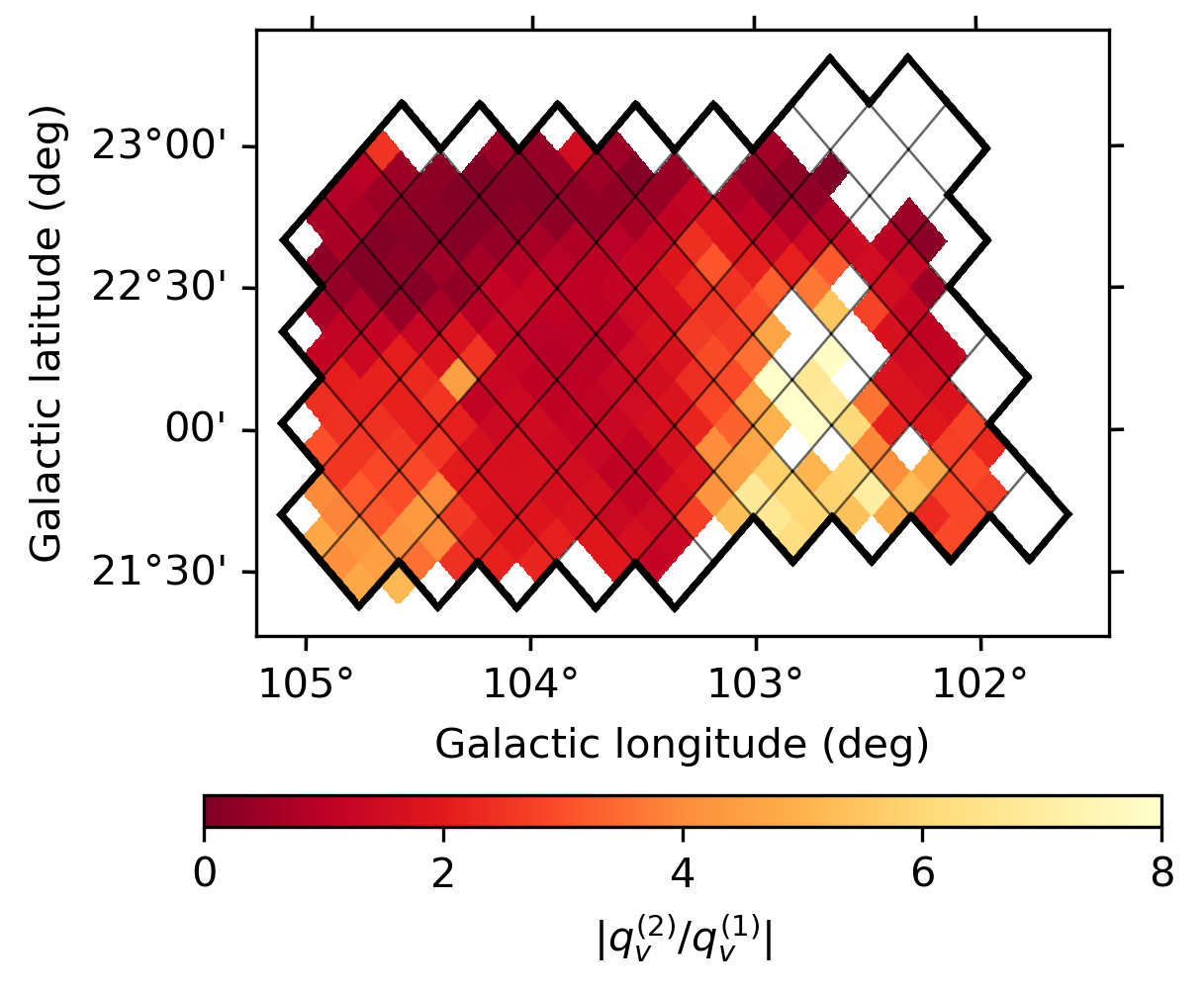}

    \caption{Map of $\lvert q^{(2)}_{v}/q^{(1)}_{v} \rvert$ at $N_{side}$=$512$
The black contour traces the $N_{side}$=$256$ footprint used in Fig.\ref{fig:gnomonic}, and thin lines mark pixel boundaries. White pixels at the outskirt of the region indicate areas with no decomposition information, 
while white pixels within the footprint indicate locations where there is no evidence for the nearby cloud.}

    \label{fig:non_linear_q}
\end{figure}

This interpretation is further supported by the tomographic decomposition of the stellar polarization data \citep{Pelgrims2024}, which shows that some of the pixels are dominated by a single cloud, and others have an equal contribution from both clouds along the LOS. For instance, we examine how the relative contribution to the stellar $q_v$ signal varies across the field using the tomographic decomposition data obtained in \citet{Pelgrims2024}. For each pixel, we obtain the mean of the posterior distribution of the Stokes $q$ of each cloud along the LOS. We then compute the absolute ratio $\lvert q^{(2)}_{v}/q^{(1)}_{v} \rvert$ in each pixel, where $q^{(1)}_{v}$ and $q^{(2)}_{v}$ are the Stokes $q_v$ of the first and second cloud along the LOS, respectively. The resulting map (Fig. \ref{fig:non_linear_q}) reveals strong gradients: in the upper-left portion $\lvert q^{(2)}_{v}/q^{(1)}_{v} \rvert \ll 1$ (cloud 2 is weak), whereas in the right
part of the map there is a region where the opposite is true as cloud 2
induces 8 times more signal in $q_v$ than cloud 1.

These spatial variations of the relative importance of each cloud are likely the origin of non-linearities in the  \( Q_s\)--\( q_v \)  relation, and are likely also the reason that we find a notable  dependence of $R_{Qq}$ on the choice of stellar sample. The two samples vary in their properties: the \(>450\,\mathrm{pc}\) sample has a mean of \(\sim14\) stars per pixel, while the \(>2000\,\mathrm{pc}\) sample has a mean of \(\sim3\) stars per pixel. In the \(>2000\,\mathrm{pc}\) case, two effects act simultaneously: (i) the number of populated pixels falls from 86 to 71 (15 are excluded), with roughly one third of these exclusions occurring in regions where cloud~2 is weak; and (ii) the stellar sampling per pixel becomes much sparser, with the mean decreasing from \(\sim14\) to \(\sim3\) stars per pixel.

We conclude that the assumption of a single linear relation between \( Q_s \) and \( q_v \) may oversimplify the actual astrophysical complexity in this region. While a linear model can still capture an average trend, care must be taken in interpreting the slope as a global physical quantity.

\end{document}